\documentclass[preprint,12pt]{elsarticle}

\usepackage{amssymb}

\usepackage[utf8]{inputenc}
\usepackage{lmodern} 
\usepackage[T1]{fontenc}
\usepackage{microtype}

\usepackage{graphicx}
\usepackage{float}
\usepackage[hypcap]{caption}
\usepackage{subcaption}
\usepackage{algorithm2e}
\usepackage{array}
\usepackage{algorithmic}

\usepackage{url}

\usepackage{textcomp}
\usepackage{booktabs}
\usepackage{array}

\usepackage{amsfonts}

\usepackage{geometry}
\usepackage{tikz}
\usetikzlibrary{trees,positioning,shapes}

\usepackage{mathtools}
\usepackage{amssymb}
\usepackage{amsthm}
\usepackage{amsmath}
\usepackage{xcolor, colortbl}
\usepackage{appendix}

\usepackage{soul}
\usepackage{pgfplots,tikz}
\usepackage[eulergreek]{sansmath}
\usepackage{listings}
\usepackage{minted}
\pgfplotsset{compat=1.18} 

\newcommand{\inblue}[1]{\textcolor{blue}{#1}}

\newcommand{\ingray}[1]{\textcolor{black}{#1}}
\newcommand{\revision}[1]{\textcolor{black}{#1}}

\usepackage[english]{babel}
\lstnewenvironment{cmd}{%
  \lstset{language=bash}%
}{}

\journal{Computer Methods in Applied Mechanics and Engineering}

\begin{document}

\begin{frontmatter}



\title{Wildfires \revision{Quasi-}Implicit Alternative-Direction Simulations using Isogeometric Finite Element Method}


\author[inst1]{Juliusz Wasieleski}

\author[inst1]{Tomasz S\l{}u\.zalec}

\author[inst1]{Maciej Woźniak}

\author[inst1]{Marcin \L{}o\'s}

\author[inst2]{\\ Andres Medina}

\author[inst2]{Paulina Sepulveda}

\author[inst3]{Albert Oliver Serra}

\author[inst5,inst6]{Eirik Valseth}

\author[inst1,inst4]{\\ Anna Paszy\'nska}

\affiliation[inst1]{organization={AGH University of Krakow, Faculty of Computer Science},
            addressline={Al. Mickiewicza 30}, 
            city={Krak\'ow},
            postcode={30-059}, 
            country={Poland}}

\affiliation[inst2]{organization={Pontificia Universidad Católica de Valparaíso, Instituto de Matemáticas},
            addressline={Avenida Brasil 2950}, 
            city={Valparaíso},
            postcode={2340001}, 
            country={Chile}}

\affiliation[inst3]{organization={The University of Las Palmas de Gran Canaria, Faculty of Computer Science and Mathematics},
            addressline={Juan de Quesada, 30}, 
            city={Las Palmas de Gran Canaria},
            postcode={35001}, 
            country={Spain}}
\affiliation[inst5]{organization={Simula Research Laboratory},
            addressline={Kristian Augusts gate 23}, 
            city={Oslo},
            postcode={0164}, 
            country={Norway}}

\affiliation[inst6]{organization={Norwegian University of Life Sciences},
            addressline={P.O. Box 5003}, 
            city={Ås},
            postcode={NO-1432}, 
            country={Norway}}

\affiliation[inst4]{organization={Jagiellonian University, Faculty of Physics, Astronomy and Applied Computer Science},
            addressline={Go\l{}ebia 26}, 
            city={Krak\'ow},
            postcode={31-007}, 
            country={Poland}}

\author[inst1]{ Maciej Paszy\'nski}

\begin{abstract}
We develop a wildfire simulation model that evolves the temperature scalar field using an energy balance equation accounting for heat generation, transport, and loss. For these equations, we develop \revision{quasi-}implicit time integration schemes using direction splitting of the differential operators. \revision{We use the Peaceman-Rachford and Strang splitting methods, including the Crank-Nicolson method}. Based on these discretizations, we derive variational formulations and explore the Kronecker product structure of the matrices.
\revision{In the wildfire model, there are some non-linear terms that we treat explicitly. We perform a detailed analysis of how treating these terms affects the stability of the time integration scheme. Namely, we show that a quasi-implicit time integration scheme achieves 10 times higher simulation accuracy.}
\revision{We present two wildfire simulations.}
The first is a simulation of the 2024 wildfire disaster in the Valparaíso region of Chile. The second one is a simulation of the 2019 wildfire disaster in Las Palmas de Gran Canaria, Spain. We discuss the numerical results and compare them against satellite images and measurement records.
\revision{We also present a numerical experiment for comparison with the state-of-the-art wildfire simulation model FARSITE.} 
Our sequential code has a linear computational cost of ${\cal O}(N)$. We also present the parallel scalability of the WILDFIRE-IGA-ADS code to illustrate the possibility of running the code on a local workstation.

\end{abstract}


\end{frontmatter}

\section{Introduction}

{Wildfire modeling has been an active research area for several decades, with approaches spanning empirical fire spread models, fully physics-based simulations, coupled atmosphere–fire systems, and, more recently, data-driven methods. These approaches differ significantly in physical fidelity, numerical formulation, and computational scalability.


There are four main groups of computational wildfire simulators available in the community.

The first group focuses on high-fidelity physical models. These models solve the Navier-Stokes equations, often using Large Eddy Simulation (LES) to model turbulence. Combustion is typically governed by a multi-step kinetic mechanism where the reaction rate follows an Arrhenius-type dependency \cite{mcgrattan2020fds}. Frameworks such as FIRETEC \cite{linn2002numerical} and WFDS \cite{mell2007numerical} provide deep insights into the physics of fire-atmosphere interaction; however, they incur high computational costs and require small time steps to satisfy the Courant-Friedrichs-Lewy condition in Computational Fluid Dynamics.

The second group of wildfire simulators is based on empirical and semi-empirical approaches. The cornerstone of operational fire management is the semi-empirical model developed by Rothermel \cite{rothermel1972mathematical}. These models bypass the micro-scale chemistry in favor of an algebraic "Rate of Spread" (ROS) formulation derived from laboratory-scale experiments. These methods are computationally efficient and are used in widely deployed systems such as FARSITE \cite{finney1998farsite}.

The third group of wildfire simulators is coupled fire-atmosphere systems and level-set methods. To bridge the gap between physical rigor and computational speed, coupled solvers like WRF-SFIRE \cite{w3} have emerged. These systems use the Level Set method to track the fire front as the zero contour of a higher-dimensional function governed by the Hamilton-Jacobi equation. This approach allows two-way coupling, in which the fire heat flux is fed back into the atmospheric model to simulate fire-induced winds.

The fourth group employs the stochastic and AI approaches. While the aforementioned deterministic models focus on the high-fidelity representation of a single fire event, the field has recently expanded into stochastic risk assessment and data-driven forecasting. Tools such as BurnP3 \cite{parisien2005burnp3,ERNI2024104221} shift the focus toward long-term planning by utilizing Monte Carlo simulations to generate burn probability maps. This approach integrates spatial fuel data with historical weather patterns to evaluate landscape-level vulnerability rather than to precisely track fronts. Furthermore, satellite-derived measurement data enables the development of machine-learning-based frameworks such as FireCast \cite{firecast2023}. These systems bypass the numerical stiffness of traditional PDE-based solvers by leveraging deep learning architectures to predict fire risk and spread based on real-time remote sensing data.

In this paper, we derive the semi-implicit wildfire simulator based on the isogeometric finite element method. It offers higher-order continuity and a level of accuracy one order of magnitude higher than a fully explicit finite element method solver.


Early operational wildfire simulators, such as perimeter- and rate-of-spread–based models, rely on empirical or semi-empirical formulations derived from laboratory experiments and field observations. While these models are computationally efficient and widely used in practice, they typically lack a consistent representation of the underlying thermodynamic and transport processes governing fire behavior. As a result, their applicability is limited when strong coupling between fire dynamics, fuel heterogeneity, and atmospheric forcing is required.

To address these limitations, coupled atmosphere–fire models were introduced, most notably the WRF-Fire framework \cite{w1,w2}. By embedding a fire spread model within the Weather Research and Forecasting (WRF) system, WRF-Fire enables two-way interactions between wildfire evolution and atmospheric dynamics, capturing effects such as fire-induced winds and plume development. Subsequent work extended this framework to real wildfire events and operational scenarios \cite{w3,w4}. Despite their physical realism, such coupled models are computationally demanding and typically require high-performance computing resources, which restricts their use in ensemble forecasting, uncertainty quantification, and faster-than-real-time simulations.

Fully physics-based computational fluid dynamics (CFD) models, including FIRETEC \cite{w5} and WFDS \cite{w6}, resolve combustion, heat transfer, and turbulent flow processes at fine spatial and temporal scales. These models provide detailed insight into fire–atmosphere interactions and fuel combustion mechanisms, but are generally limited to small domains and short simulation times due to their high computational cost. Consequently, their use is primarily confined to fundamental studies and validation experiments rather than large-scale predictive modeling.

More recently, reduced-order and data-driven approaches have been explored to improve computational efficiency. Physics-informed neural networks (PINNs) approximate the solution of governing partial differential equations by embedding physical constraints into neural network training \cite{w7}. \revision{This is illustrated in the FARSITE physics-informed neural network solver \cite{w8}}. However, these methods rely on optimization-based training procedures, may suffer from convergence and stability issues, and often exhibit limited robustness in the presence of sharp fronts, stiff reaction terms, and heterogeneous input data.

In this work, we consider a reduced wildfire energy balance model and focus on its numerical approximation using B-spline finite element discretizations within an isogeometric analysis framework \cite{HUGHES20054135}. The spatial discretization is constructed using tensor-product B-spline finite element spaces, which provide higher-order continuity and improved approximation properties compared to standard $C^0$ Lagrange finite elements. B-spline basis functions of degree $p$ exhibit $C^{p-1}$ continuity across element interfaces, leading to enhanced accuracy per degree of freedom, particularly for diffusion-dominated operators. The resulting smooth discrete temperature fields are well-suited for modeling thermal diffusion and radiative heat transfer, which involve second-order spatial derivatives and nonlocal effects.

This wildfire simulator is an extension of the explicit solver performing a sequence of $L^2$ projections, employed for simulations of tumor growth in 2D and 3D \cite{LOS20171257,LOS20191}, atmospheric simulations \cite{LOS2024102238}, electromagnetic wave generations \cite{LOS202336}, among others.
We selected B-splines for the easy construction of tensor-product grids, but there are several other options for discretization available \cite{10.1145/882262.882295,10.1145/1015706.1015715,SCOTT2012206,WEI2017349,
VUONG20113554,BORNEMANN2013584,GIANNELLI2012485,DOKKEN2013331,
JOHANNESSEN2014471,https://doi.org/10.1111/j.1467-8659.2010.01766.x,
WEI20151}.

From an analytical perspective, the increased inter-element continuity of B-spline spaces yields discrete operators with favorable spectral properties, including reduced numerical dispersion and improved stability under implicit time integration. These characteristics are especially advantageous in wildfire simulations, where sharp temperature gradients coexist with large smooth regions and where numerical robustness under strong wind-driven transport is essential. Moreover, the tensor-product structure of the B-spline spaces on structured grids naturally induces separable mass, stiffness, and advection operators, enabling Kronecker product representations of the discrete operators.

Time integration is performed using \revision{quasi-}implicit operator-splitting techniques designed to exploit this tensor-product structure. We investigate two classes of schemes: an alternating-direction Peaceman–Rachford method and symmetric Strang splitting schemes \revision{combined with Crank–Nicolson time integration}. Both approaches rely on the same directional decomposition of the linear transport–diffusion operator, enabling a consistent comparison of stability, accuracy, and computational efficiency within a unified isogeometric framework.

A key advantage of the proposed methods is that the resulting fully discrete systems inherit a Kronecker product structure from the tensor-product B-spline basis. This allows the multidimensional transport–diffusion problem to be reduced to a sequence of one-dimensional solves at each time step, significantly reducing computational cost and memory requirements compared to fully coupled solvers.

Nonlinear source terms associated with combustion, radiation, and convective heat exchange are treated explicitly or semi-implicitly at intermediate time levels. This strategy avoids fully nonlinear systems while preserving the temporal accuracy of the splitting schemes.

The Peaceman–Rachford alternating direction scheme provides a robust, first-order-accurate splitting for stiff transport–diffusion operators. The Strang splitting schemes combined with Crank–Nicolson discretizations further enhance accuracy while retaining robustness. The Crank–Nicolson variant achieves second-order temporal accuracy in transport-dominated regimes.

\revision{We provide a detailed numerical investigation on the quasi-implicit time integration scheme, starting from the linear advection-diffusion equations with separable coefficients, and incorporating particular non-linear terms. We measure numerical accuracy as a function of the time step using the manufactured solution technique.}

The novelties of this paper are the following.

\begin{itemize}
\item The presented approach combines isogeometric spatial discretization with alternating-direction implicit time integration schemes for wildfire simulation. Our previous work included the development of these schemes for advection-diffusion equations \cite{LOS2020213} as well as for time-dependent Stokes and Navier-Stokes equations \cite{los2020isogeometric}. \revision{In this work, we investigate how incorporating the non-linear terms and treating them explicitly influences the convergence of the method}.

\item \revision{We provide a C++ open-source code for the wildfire simulations, and include the installation and usage manuals. The code extends the IGA-ADS library \cite{LOS201799}.}

\item \revision{We perform simulations of the two wildfires and compare with measurement data.
The first concerns the Las Palmas de Gran Canaria 2019 disaster, and the second concerns the Valparaíso region of Chile 2024 disaster.}
\end{itemize}

The structure of the paper is the following. First, the wildfire model is derived in Section 2. Next, Section 3 describes the time integration schemes, including the Peaceman-Rachford time integration in Section 3.1 and
Strang splitting schemes, with Crank-Nicolson methods. The derivations include variational formulations, discretizations with B-spline basis functions, and the derivation of Kronecker product matrices. Next, Section 4 estimates the computational complexity per time step. The next Section 5 describes the IGA-ADS-WILDFIRE code as implemented in the IGA-ADS environment. In the remaining part of the paper, the numerical experiments are presented in Section 6, including comparisons with the FAIRSITE simulator, estimates of stability, convergence, and the order of the time integration schemes, as well as simulations of the Chilean and Spanish wildfire disasters. Section 7 is devoted to scalability experiments of the shared-memory IGA-ADS-WILDFIRE implementation. We conclude the paper in Section 8.

\section{Wildfire model}

We consider a reduced wildfire energy balance model originally introduced in
\cite{w8}, with targeted modifications aimed at improving numerical
robustness and compatibility with \revision{quasi-}implicit operator-splitting and
alternating-direction time integration schemes. 

The formulation adopted here balances physical fidelity with computational
tractability, making it suitable for large-scale numerical simulations while
retaining the dominant mechanisms governing wildfire dynamics.

\subsection{Governing equations}

The evolution of the temperature field $T(\boldsymbol{x},t)$ is governed by an
energy balance equation accounting for heat generation, transport, and loss
mechanisms:
\begin{equation}
\rho \frac{\partial}{\partial t}\left(c_p T\right)
=
R_C + Q_W
- \nabla \cdot \left(q_C + q_D + q_r\right)
+ Q_{\mathrm{conv}} + Q_{rz},
\label{eq:energy}
\end{equation}
where $c_p$ denotes the specific heat capacity at constant pressure and $\rho$ is
the density of the gas mixture. Equation \eqref{eq:energy} consists of a linear
transport--diffusion operator acting on the temperature field, coupled with
nonlinear source and loss terms arising from combustion, radiation, and convective
exchange. From a mathematical perspective, the resulting temperature equation is a nonlinear
parabolic advection--diffusion--reaction problem with temperature-dependent
coefficients.

\paragraph{Combustion heat release}
The volumetric heat release due to combustion is modeled following
\cite{w8} as
\begin{equation}
R_C = \rho\, c_h\, h_c\, \frac{M}{M_1}\, r,
\end{equation}
where $h_c<0$ denotes the specific combustion enthalpy, $c_h$ is an enthalpy
correction coefficient, $M$ is the molar mass of the gas mixture, and $M_1$ is the
molar mass of the fuel. With this convention, $R_C$ acts as a positive heat source
in the energy balance. The chemical reaction rate $r$ depends on the local
temperature and on the availability of fuel and oxidizer.

\paragraph{Wind-driven heat transport}
The contribution of atmospheric wind to heat transport is modeled by
\begin{equation}
Q_W = -\rho c_W\, \boldsymbol{b}\cdot\nabla\left(c_p T\right),
\end{equation}
where $\boldsymbol{b}=(b_x,b_y)$ is a prescribed wind velocity field and $c_W$ is a
reduction coefficient accounting for vegetation drag and subgrid-scale effects. This
term introduces a first-order advective transport operator into the governing
equation.

\paragraph{Conductive and diffusive heat fluxes}
Thermal conduction is described by Fourier’s law,
\begin{equation}
q_C = -\kappa \nabla T,
\end{equation}
where $\kappa$ denotes an effective thermal conductivity. The interdiffusion
enthalpy flux $q_D$ arises from species diffusion and is formally given by
\begin{equation}
q_D = \sum_i h_i J_i,
\end{equation}
with $J_i$ denoting the diffusive mass flux of species $i$. In the numerical
experiments presented in this work, this term is neglected for simplicity and to
avoid additional coupling between the temperature and species equations, without
altering the structure of the numerical schemes.

\paragraph{Radiative heat transfer}
Radiative heat transport within the vegetation layer is modeled as a nonlinear
diffusive flux,
\begin{equation}
q_r = -4\sigma\epsilon\delta_x T^3 \nabla T,
\end{equation}
where $\sigma$ is the Stefan--Boltzmann constant, $\epsilon$ is the surface
emissivity, and $\delta_x$ is an effective absorption length. This term introduces a
temperature-dependent diffusion coefficient and constitutes a major source of
nonlinearity in the model.

\paragraph{Convective and vertical radiative losses}
Heat exchange with the ambient atmosphere is represented by a Newton cooling term,
\begin{equation}
Q_{\mathrm{conv}} = \chi\left(T_{\mathrm{amb}} - T\right),
\end{equation}
where $\chi$ is the convective heat transfer coefficient and $T_{\mathrm{amb}}$ is
the ambient temperature. Vertical radiative losses are modeled by
\begin{equation}
Q_{rz} = \sigma\epsilon\delta_z^{-1}\left(T_{\mathrm{amb}}^4 - T^4\right), \label{Tamb}
\end{equation}
where $\delta_z$ denotes an effective vertical emission length scale.

\paragraph{Combustion modeling}
The detailed evolution of chemical species is not resolved explicitly in the present
formulation. Instead, the net effect of fuel consumption and chemical reactions is
incorporated into the effective combustion heat release term $R_C$. This reduced
modeling approach is consistent with the objective of capturing the dominant thermal
dynamics of wildfire propagation while avoiding the introduction of additional
unknowns and stiffness associated with multi-species reaction kinetics.

\paragraph{Fuel availability coefficient}
Fuel heterogeneity is modeled through a spatially varying coefficient
$\eta(x)\in[0,1]$, representing the local availability of combustible material.
This coefficient is constructed offline from satellite-based NDVI data combined
with a land-cover classification mask (WorldCover), which excludes non-combustible
regions such as urban areas, water bodies, and bare soil.
The resulting continuous field is stored as a rasterized CSV file and
interpolated onto the finite element mesh using B-spline basis functions.

\revision{In particular, the amount of green color on satellite images indicates the forest, which is a potentially flammable element that is the main source of the spreading fire. 
We extract the green channel from the satellite image using the presented Python code. }
\begin{figure}
    \centering
    \includegraphics[width=0.48\textwidth]{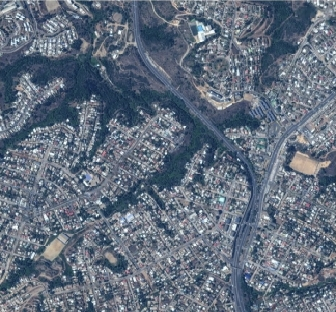}    \includegraphics[width=0.45\textwidth]{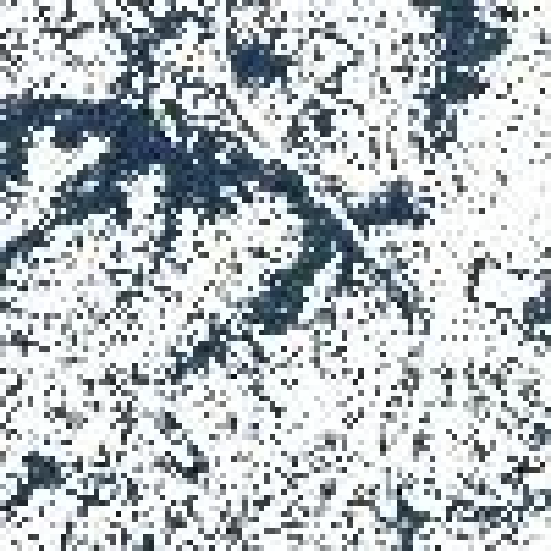}
   \caption{\revision{Satellite image on the left and the fuel map on the right for the Chile wildfire simulation. }}
    \label{fig:buttle}
\end{figure}
\begin{lstlisting}[
    language=Python,
    caption={Fuel map extraction.},
    basicstyle=\ttfamily,
    keywordstyle=\color{blue},
    stringstyle=\color{red},
    commentstyle=\color{green},
    frame=single,
    numbers=left,
    numberstyle=\tiny\color{gray},
    stepnumber=1,
    numbersep=5pt,
    backgroundcolor=\color{lightgray!25}
]
import cv2
import matplotlib.pyplot as plt
import pandas as pd
im = cv2.imread(''forest_map.png'')
imbw = im[:, :, 1]  # only green channel
imbw = imbw.reshape(imbw.shape[0], imbw.shape[1])
imbw = imbw / 255
df = pd.DataFrame(imbw)
df.to_csv(''image .csv '', index=False, header=False)
\end{lstlisting}
\revision{For example, the satellite image of the Valparaiso area of Chile and the fuel map extracted by the presented python code is illustrated in Figure \ref{fig:buttle}.}

At the discrete level, the fuel coefficient enters the numerical scheme by
scaling the combustion reaction rate,
\[
r(x,T) \;\longrightarrow\; \eta(x)\, r(T),
\]
thereby modulating the local heat release term in the energy balance equation.
This approach preserves the structure of the governing equations while enabling realistic spatial control of fire propagation.

\paragraph{Vegetation heterogeneity and NDVI-based fuel availability}
Spatial heterogeneity of the fuel load is incorporated through a vegetation
availability map derived from satellite-based Normalized Difference Vegetation
Index (NDVI) data. The NDVI field is transformed into a dimensionless coefficient
$P_{\mathrm{NDVI}}(x)\in[0,1]$, which represents the local probability of ignition
and the effective availability of combustible material. Regions with sparse or
absent vegetation correspond to values close to zero, while densely vegetated
areas attain values close to one.

In the reduced modeling framework adopted here, vegetation heterogeneity affects
only the local heat generation due to combustion, without modifying the transport
or diffusion mechanisms. Accordingly, the NDVI-based coefficient multiplies the
combustion source term in the energy balance equation, while all other terms remain
unchanged.

The energy balance equation \eqref{eq:energy} is therefore modified as
\begin{equation}
\rho \frac{\partial}{\partial t}\left(c_p T\right)
=
P_{\mathrm{NDVI}}(x)\, R_C
+ Q_W
- \nabla \cdot \left(q_C + q_D + q_r\right)
+ Q_{\mathrm{conv}} + Q_{rz}.
\label{eq:energy_ndvi}
\end{equation}
The coefficient $P_{\mathrm{NDVI}}(x)$ is obtained by normalizing the NDVI field
according to
\begin{equation}
P_{\mathrm{NDVI}}(x)
=
\frac{\mathrm{NDVI}(x)-\mathrm{NDVI}_{\min}}
{\mathrm{NDVI}_{\max}-\mathrm{NDVI}_{\min}},
\end{equation}
followed by thresholding to enforce values in the interval $[0,1]$.

From a numerical standpoint, $P_{\mathrm{NDVI}}(x)$ is a prescribed, time-independent
coefficient and enters the formulation exclusively through the explicit source
term. As a consequence, it does not alter the structure of the transport--diffusion
operator nor the unconditional stability properties of the implicit splitting
schemes considered in this work.

\paragraph{Ignition initialization}
The ignition region is introduced exclusively through the initial condition of
the temperature field.
A spatial ignition mask $\gamma(x)\in[0,1]$ is constructed from the reported
geographical ignition point of the wildfire and smoothed using a Gaussian
distribution to account for subgrid-scale uncertainty.

\paragraph{Boundary conditions}
The computational domain $\Omega \subset \mathbb{R}^2$ is equipped with homogeneous
Neumann boundary conditions for all horizontal heat fluxes. More precisely, we
impose
\[
(q_C + q_r)\cdot \mathbf{n} = 0 \qquad \text{on } \partial\Omega,
\]
where $\mathbf{n}$ denotes the outward unit normal vector.
This condition corresponds to a vanishing net lateral heat flux across the domain
boundary and is naturally enforced by the variational formulation.

Vertical heat exchange with the atmosphere is modeled exclusively through the
source terms $Q_{\mathrm{conv}}$ and $Q_{rz}$ in the energy balance equation.
No Dirichlet boundary conditions are prescribed for the temperature field, as this
would introduce artificial cooling effects and compromise energy consistency.

\subsection{Model parameters}

The quantities appearing in equations \eqref{eq:energy}–\eqref{Tamb} depend on
a number of physical and empirical parameters. Several of these parameters,
particularly those associated with combustion kinetics and radiative heat transfer,
are effective or calibrated quantities commonly employed in reduced wildfire
models. Their values are adopted from \cite{w8,Varadachari2009} and are
chosen to represent vegetation-scale combustion processes, heat transfer
mechanisms, and atmosphere--fuel interactions that are not explicitly resolved by
the present formulation.

Unless otherwise stated, parameter values are scaled or normalized consistently
with the reduced modeling framework and should be interpreted as effective model
coefficients rather than laboratory-measured thermodynamic constants. The complete
set of parameters used in the numerical simulations is summarized in
Table~\ref{tab:parameters}.

\begin{table}[h]
\centering
\renewcommand{\arraystretch}{1.15}
\begin{tabular}{lll}
\hline
\textbf{Symbol} & \textbf{Description} & \textbf{Value / Units} \\
\hline
$c_p$ & Specific heat capacity & $1.0\;\mathrm{J\,kg^{-1}K^{-1}}$ \\
$\rho$ & Gas density & $1.293\;\mathrm{kg\,m^{-3}}$ \\
$\kappa$ & Thermal conductivity & $0.3\;\mathrm{W\,m^{-1}K^{-1}}$ \\
$\sigma$ & Stefan--Boltzmann constant & $5.67\times10^{-8}\;\mathrm{W\,m^{-2}K^{-4}}$ \\
$\epsilon$ & Emissivity & $0.05$ \\
$c_h$ & Enthalpy correction coefficient & $1.0$ \\
$h_c$ & Combustion enthalpy & $-70\;\mathrm{J\,kg^{-1}}$ \\
$c_W$ & Wind reduction coefficient & $0.5$ \\
$\boldsymbol{b}$ & Wind velocity field & $\mathrm{m\,s^{-1}}$ \\
$\chi$ & Convective coefficient & $2\times10^{-2}\;\mathrm{W\,m^{-2}K^{-1}}$ \\
$T_{\mathrm{amb}}$ & Ambient temperature & $300\;\mathrm{K}$ \\
$T_{ig}$ & Ignition temperature & $800\;\mathrm{K}$ \\
$\delta_x$ & Radiative absorption length & $3.5\times10^{-2}/\epsilon\;\mathrm{m}$ \\
$\delta_z$ & Vertical emission length & $1.5\,\epsilon\;\mathrm{m}$ \\
\hline
\end{tabular}
\caption{Model parameters used in the wildfire simulations.}
\label{tab:parameters}
\end{table}

\section{Numerical discretizations}

This section is devoted to the numerical approximation of the wildfire energy balance model introduced in the previous section. The spatial discretization is carried out using tensor-product B-spline finite element spaces
within an isogeometric analysis framework, which naturally induces separable mass, stiffness, and advection operators. This tensor-product structure is exploited by \revision{Peaceman-Rachford} and Strang splitting schemes, allowing the multidimensional problem to be reduced to a sequence of one-dimensional solves while
retaining accuracy and robustness for large-scale wildfire simulations.

\revision{Substitution of the model parameters into the wildfire equations results in the following general terms
\begin{eqnarray}
\underbrace{ \frac{\partial T}{\partial t}}_{\textrm{\textrm{time progression}}} + \inblue{\underbrace{C_{\textrm{advection}}  \mathbf{b} \cdot \nabla T}_{\textrm{advection}} -\underbrace{C_{\textrm{diffusion}} \nabla \cdot  \nabla T}_{\textrm{diffusion}} -\underbrace{C_{\textrm{reaction}}   T}_{\textrm{reaction}} } 
 \notag \\
-\ingray{\underbrace{\nabla \cdot (C_{\textrm{non-linear diffusion}} \cdot T^3 \nabla T)}_{\textrm{non-linear diffusion}}} =
\notag \\ \ingray{ C_{\textrm{ignition}} \cdot [T>T_{ig} \textrm{ AND fuel}>0.2]T \exp(\frac{-300}{T})+}   \notag
\\  \ingray{\underbrace{ \qquad \qquad \qquad \qquad \qquad + C_{\textrm{forcing}}- C_{\textrm{radiation}} \cdot  T^4}_{\textrm{non-linear forcing}}}
\notag
\end{eqnarray}
The time progression and advection-diffusion-reaction terms fit exactly into the implicit time integration scheme, preserving the Kronecker product structure of the matrices. The remaining terms are non-linear and non-separable. They involve the non-linear diffusion term of $\nabla \cdot \left( T^3 \nabla T\right)$, the non-linear term containing the ignition condition (depending on the fuel concentration and ignition temperature) multiplied by the exponential term of $\exp(\frac{-300}{T})$, and the non-linear radiative term of $T^4$. In the following sections,, we will investigate the convergence of the method when including these nonlinear terms on the right-hand side.}

\subsection{Peaceman--Rachford ADI scheme}

The wildfire temperature equation exhibits strong stiffness arising from diffusive, radiative, and combustion-induced source terms. We employ an implicit alternating-direction scheme of Peaceman-Rachford (PR) type. This approach allows the dominant transport-diffusion operators to be treated implicitly while decoupling the multidimensional problem into a sequence of one-dimensional solves.

\revision{The time discretization based on the Peaceman-Rachford scheme reads
\begin{equation}\label{PR}
\displaystyle{\left\{
\begin{split}
\frac{T^{n+1/2}-T^{n}}{\tau/2}+\mathcal{L}_{1}T^{n+1/2}&=F^{n+1/2}-\mathcal{L}_{2}T^{n}, \nonumber \\
\frac{T^{n+1}-T^{n+1/2}}{\tau/2}+\mathcal{L}_{2}T^{n+1}&=F^{n+1/2}-\mathcal{L}_{1}T^{n+1/2}. \nonumber \\
\end{split}
\right.} 
\end{equation}
where
\begin{equation}\label{termsPR}
\begin{aligned}
\mathcal{L}_{1}T =& C_{\textrm{advection}}b_x \partial_x T - C_{\textrm{diffusion}}  \partial_{xx} T -C_{\textrm{reaction}}   T \notag \\
\mathcal{L}_{2}T =& C_{\textrm{advection}}b_y \partial_x T - C_{\textrm{diffusion}}  \partial_{yy} T \notag \\
F = &
 C_{\textrm{non-linear diffusion}} \nabla \cdot ( \cdot T^3 \nabla T) + \notag \\ & C_{\textrm{ignition}} \cdot [T>T_{ig} \textrm{ AND fuel}>0.2]T \exp(\frac{-300}{T}) + C_{\textrm{forcing}}- C_{\textrm{radiation}} \cdot  T^4 \notag
 \end{aligned}
\end{equation}}
Here $\mathcal{L}$ denotes the linear transport--diffusion operator and $F$ collects
the nonlinear source and loss terms, including combustion heat release, convective
exchange with the ambient atmosphere, and vertical radiative losses.
The remaining source terms are grouped into the forcing term $F$.

All nonlinear source terms, including combustion heat release, convective losses, and
vertical radiative losses, are evaluated explicitly or semi-implicitly at the
intermediate time level $t^{n+1/2}$.

\paragraph{Variational formulation}
Let $V_h \subset H^1(\Omega)$ be a conforming finite-dimensional space. The variational
form of \eqref{PR} reads: find $T^{n+1/2}, T^{n+1} \in V_h$ such that for all
$v \in V_h$,
\revision{
\begin{equation} 
\left\{
\begin{aligned}
(T^{n+1/2},v)
&+ \frac{\tau}{2}\left(C_{\textrm{diffusion}} \frac{\partial T^{n+1/2}}{\partial x},
\frac{\partial v}{\partial x}\right)
+ \frac{\tau}{2}\left(C_{\textrm{advection}} \frac{\partial T^{n+1/2}}{\partial x}, v \right) + \\
&- \frac{\tau}{2}\left( C_{\textrm{reaction}} T^{n+1/2},v \right)=
(T^{n},v)
- \frac{\tau}{2}\left(C_{\textrm{diffusion}} \frac{\partial T^{n}}{\partial y},
\frac{\partial v}{\partial y}\right)+\\
&- \frac{\tau}{2}\left(C_{\textrm{advection}} \frac{\partial  T^{n}}{\partial y}, v \right)
+ \frac{\tau}{2}(F^{n+1/2},v),
\\
(T^{n+1},v)
&+ \frac{\tau}{2}\left(C_{\textrm{diffusion}} \frac{\partial T^{n+1}}{\partial y},
\frac{\partial v}{\partial y}\right)
+ \frac{\tau}{2}\left(C_{\textrm{advection}} \frac{\partial T^{n+1}}{\partial y}, v \right) =
(T^{n+1/2},v)+ \\
& - \frac{\tau}{2}\left(C_{\textrm{diffusion}} \frac{\partial T^{n+1/2}}{\partial x},
\frac{\partial v}{\partial x}\right)
- \frac{\tau}{2}\left( C_{\textrm{advection}}  \frac{\partial  T^{n+1/2}}{\partial x}, v \right)+\\
&+ \frac{\tau}{2}\left( C_{\textrm{reaction}} T^{n+1/2},v \right)
+ \frac{\tau}{2}(F^{n+1/2},v).
\end{aligned}
\right. \nonumber
\label{eq:varPR_wildfire}
\end{equation}}
For clarity of the presentation, boundary conditions are omitted and assumed to be
incorporated consistently within the variational formulation.

\paragraph{Matrix formulation}
Let $\Omega = \Omega_x \times \Omega_y$ be a rectangular domain and
$V_h = V_h^x \otimes V_h^y$ a tensor-product finite element or B-spline space. 
\revision{We define $M^{x,y}$, $K^{x,y}$ and $G^{x,y}$ the 1D mass, stiffness and advection matrices, respectively, as
\begin{equation}
\{ M^s \}_{i,k} =  \int_{\Omega_s} B_i B_k \,\mbox{d}s,  \;
 \{ S^s \}_{i,k} =  \int_{\Omega_s} \frac{\partial B_i}{\partial s} \frac{\partial B_k}{\partial s} \,\mbox{d}s, \; 
\{ G^s \}_{i,k} =  \int_{\Omega_s} \frac{\partial B_i}{\partial s}  B_k \,\mbox{d}s
\end{equation}
with $s \in \{x,y\}$. We do not explicitly write the implementation constants, but we assume they are hidden within the matrix definitions for clarity of presentation.} The Peaceman--Rachford scheme can be written in Kronecker product form as
\begin{equation}
\left\{
\begin{aligned}
&\Bigl[M^x + \frac{\tau}{2}(K^x + G^x)\Bigr]
\otimes M^y \;\mathbf{T}^{n+1/2}
=
M^x \otimes
\Bigl[M^y - \frac{\tau}{2}(K^y + G^y)\Bigr]
\mathbf{T}^{n}
+ \frac{\tau}{2}\mathbf{F}^{n+1/2},
\\[1ex]
&M^x \otimes
\Bigl[M^y + \frac{\tau}{2}(K^y + G^y)\Bigr]
\mathbf{T}^{n+1}
=
\Bigl[M^x - \frac{\tau}{2}(K^x + G^x)\Bigr]
\otimes M^y \;\mathbf{T}^{n+1/2}
+ \frac{\tau}{2}\mathbf{F}^{n+1/2}.
\end{aligned}
\right.
\label{eq:matrixPR}
\end{equation}
Each time step, therefore, reduces to a sequence of one-dimensional linear solves in the $x$- and $y$-directions, significantly reducing the computational cost.

\subsection{\revision{Strang splitting scheme with Crank-Nicolson method}}

\revision{In the Strang splitting scheme we divide problem $T_{t}+\mathcal{L}T=F$ into 
\begin{equation}\label{subStrang}
\displaystyle{\left\{
\begin{split}
P_{1}:&\;\partial_{t} T+\mathcal{L}_{1}T=F,\\
P_{2}:&\;\partial_{t} T+\mathcal{L}_{2}T=0,\\
\end{split}
\right.} \nonumber 
\end{equation}
the scheme integrates the solution from $T_{n}$ to $T_{n+1}$ into substeps:
\begin{equation}\label{schemeStrang}
\displaystyle{\left\{
\begin{split}
\mbox{Solve}\;P_{1}:&\;\partial_{t} T+\mathcal{L}_{1}T=F,\;\mbox{in}\;(t_{n},t_{n+1/2}),\\
\mbox{Solve}\;P_{2}:&\;\partial_{t} T+\mathcal{L}_{2}T=0,\;\mbox{in}\;(t_{n},t_{n+1}),\\
\mbox{Solve}\;P_{1}:&\;\partial_{t} T+\mathcal{L}_{1}T=F,\;\mbox{in}\;(t_{n+1/2},t_{n+1}),\\
\end{split}
\right.} \nonumber 
\end{equation}.}
where again
\begin{equation}\label{termsPR}
\begin{aligned}
\mathcal{L}_{1}T =& C_{\textrm{advection}}b_x \partial_x T - C_{\textrm{diffusion}}  \partial_{xx} T -C_{\textrm{reaction}}   T \notag \\
\mathcal{L}_{2}T =& C_{\textrm{advection}}b_y \partial_x T - C_{\textrm{diffusion}}  \partial_{yy} T \notag \\
F = &
 C_{\textrm{non-linear diffusion}} \nabla \cdot ( \cdot T^3 \nabla T) + \notag \\ & C_{\textrm{ignition}} \cdot [T>T_{ig} \textrm{ AND fuel}>0.2]T \exp(\frac{-300}{T}) + C_{\textrm{forcing}}- C_{\textrm{radiation}} \cdot  T^4 \notag
 \end{aligned}
\end{equation}}

To improve temporal accuracy, we next apply the Crank--Nicolson method to each
subproblem of the Strang splitting scheme. The resulting time discretization reads
\revision{\begin{equation}\label{CN}
\displaystyle{\left\{
\begin{split}
&\frac{T^{n+1/2}-T^{n}}{\tau/2}+\frac{1}{2}(\mathcal{L}_{1}T^{n+1/2}+\mathcal{L}_{1}T^{n})=\frac{1}{2}(F^{n+1/2}+F^{n}), \nonumber\\
&\frac{T^{n+1}-T^{n}}{\tau}+\frac{1}{2}(\mathcal{L}_{2}T^{n+1}+\mathcal{L}_{2}T^{n})=0, \nonumber \\
&\frac{T^{n+1}-T^{n+1/2}}{\tau/2}+\frac{1}{2}(\mathcal{L}_{1}T^{n+1}+\mathcal{L}_{1}T^{n+1/2})=\frac{1}{2}(F^{n+1}+F^{n+1/2}). \nonumber \\
\end{split}
\right.} \nonumber
\end{equation}}

\paragraph{Variational formulation}
The corresponding weak formulation is
\revision{\begin{equation}\label{varCN}
\displaystyle{\left\{
\begin{split}
(T^{n+1/2},v)&+\frac{\tau}{4}\left(C_{\textrm{diffusion}}\frac{\partial T^{n+1/2}}{\partial x},\frac{\partial v}{\partial x}\right)+\frac{\tau}{4}\left(C_{\textrm{advection}}b_{x}\frac{\partial T^{n+1/2}}{\partial x},v\right)+ \\
&-\frac{\tau}{4}\left(C_{\textrm{reaction}}T^{n+1/2},v\right)=
(T^{n},v)-\frac{\tau}{4}\left(C_{\textrm{diffusion}}\frac{\partial T^{n}}{\partial x},\frac{\partial v}{\partial x}\right)+\\ & -\frac{\tau}{4}\left(C_{\textrm{advection}}b_{x}\frac{\partial T^{n}}{\partial x},v\right)+\frac{\tau}{4}\left(C_{\textrm{reaction}}T^{n},v\right)+\frac{\tau}{4}(F^{n+1/2}+F^{n},v),
\end{split}
\right.} \nonumber
\end{equation}}
\revision{\begin{equation}\label{varCN}
\displaystyle{\left\{
\begin{split}
(T^{n+1},v)&+\frac{\tau}{2}\left(C_{\textrm{diffusion}}\frac{\partial T^{n+1}}{\partial y},\frac{\partial v}{\partial y}\right)+\frac{\tau}{2}\left(C_{\textrm{advection}}b_{y}\frac{\partial T^{n+1}}{\partial y},v\right)=\\
&=(T^{n},v)-\frac{\tau}{2}\left(C_{\textrm{diffusion}}\frac{\partial T^{n}}{\partial y},\frac{\partial v}{\partial y}\right)-\frac{\tau}{2}\left(C_{\textrm{advection}}b_{y}\frac{\partial T^{n}}{\partial y},v\right),\\
\end{split}
\right.} \nonumber
\end{equation}}
\revision{\begin{equation}\label{varCN}
\displaystyle{\left\{
\begin{split}
(T^{n+1},v)&+\frac{\tau}{4}\left(C_{\textrm{diffusion}}\frac{\partial T^{n+1}}{\partial x},\frac{\partial v}{\partial x}\right)+\frac{\tau}{4}\left(C_{\textrm{advection}}b_{x}\frac{\partial T^{n+1}}{\partial x},v\right)+\\
&-\frac{\tau}{4}\left(C_{\textrm{reaction}}T^{n+1},v\right)=(T^{n+1/2},v)-\frac{\tau}{4}\left(C_{\textrm{diffusion}}\frac{\partial T^{n+1/2}}{\partial x},\frac{\partial v}{\partial x}\right)+\\&-\frac{\tau}{4}\left(C_{\textrm{advection}}b_{x}\frac{\partial T^{n+1/2}}{\partial x},v\right)+ \frac{\tau}{4}\left(C_{\textrm{reaction}}T^{n+1/2},v\right)+\frac{\tau}{4}(F^{n+1}+F^{n+1/2},v).
\end{split}
\right.} \nonumber
\end{equation}}

\paragraph{Matrix formulation}
The corresponding Kronecker product representation reads
\revision{\begin{equation}\label{matrixCN}
\displaystyle{\left\{
\begin{split}
&\left[M^{x}+\frac{\tau}{4}(K^{x}+G^{x})\right]\otimes M^{y}T^{*} = \\ & \quad \left[M^{x}-\frac{\tau}{4}(K^{x}+G^{x})\right]\otimes M^{y}T^{n}+\frac{\tau}{4}(F^{n+1/2}+F^{n}),\\
&M^{x}\otimes\left[M^{y}+\frac{\tau}{2}(K^{y}+G^{y})\right]T^{**}=M^{x}\otimes\left[M^{y}-\frac{\tau}{2}(K^{y}+G^{y})\right]T^{*},\\
&\left[M^{x}+\frac{\tau}{4}(K^{x}+G^{x})\right]\otimes M^{y}T^{n+1}= \\ & \quad \left[M^{x}-\frac{\tau}{4}(K^{x}+G^{x})\right]\otimes M^{y}T^{**}+\frac{\tau}{4}(F^{n+1}+F^{n+1/2}).\\
\end{split}
\right.} \nonumber
\end{equation}}
\revision{Here again we have assumed that the implementation constants are hiden inside the matrix defintions for the clarity of presentation.}

The Crank--Nicolson Strang splitting scheme is second-order accurate in time \revision{for the pure linear advection-diffusion-reaction problems} and
provides improved accuracy in transport-dominated regimes, while retaining the
favorable stability properties associated with the implicit treatment of the
directional transport--diffusion operators.

\section{Computational complexity}
\label{sec:complexity}

We analyze the computational complexity of the proposed operator-splitting schemes
combined with tensor-product B-spline finite element discretizations.
The key feature exploited throughout is the Kronecker product structure of the
discrete operators induced by the tensor-product B-spline basis on structured grids.
This structure enables an efficient reduction of the multidimensional transport--diffusion
problem to a sequence of one-dimensional implicit solves at each time step.

Let $N_x$ and $N_y$ denote the number of degrees of freedom in the $x$- and $y$-directions,
respectively, and let $N = N_x N_y$ be the total number of spatial degrees of freedom.
The one-dimensional mass, diffusion, and advection matrices $M^{x,y}$, $K^{x,y}$,
and $G^{x,y}$ are banded with bandwidth proportional to the B-spline degree $p$.

\paragraph{Cost of directional implicit solves}
\revision{Each substep in the Peaceman--Rachford, Strang, or Minev-Guarmound splitting schemes involves
the solution of linear systems of the form
\[
\left(M^{x} + \gamma K^{x} + \delta G^{x} \right) \otimes M^{y}
\quad \text{or} \quad
M^{x} \otimes \left(M^{y} + \gamma K^{x} + \delta G^{y}\right),
\]
where $K^{x,y}$ denotes the one-dimensional stiffness operator, $G^{x,y}$ denotes the one-dimensional advection--diffusion operator and
$\gamma \geq 0$, $\delta \geq 0$ depends on the time-step size and the chosen time integration scheme.}
By exploiting the Kronecker product structure, these systems can be solved by
a sequence of independent one-dimensional linear systems.

Specifically, each directional implicit solve requires:
\begin{itemize}
\item $N_y$ independent linear solves of size $N_x$ for $x$-direction substeps, or
\item $N_x$ independent linear solves of size $N_y$ for $y$-direction substeps.
\end{itemize}
Since the one-dimensional matrices are banded, each solve can be performed in
$\mathcal{O}(p^2 N_x)$ or $\mathcal{O}(p^2 N_y)$ operations using direct solvers.
Consequently, the overall cost of a single directional substep scales as
\[
\mathcal{O}\big(p^2 N\big).
\]

\paragraph{Cost per time step}
For the Peaceman–Rachford scheme, each time step consists of two directional
solves, resulting in a total computational cost of
\[
\mathcal{O}\big(p^2 N\big)
\quad \text{per time step}.
\]
For the Strang splitting schemes, each time step involves three directional
solves (two half steps and one full step), leading to the same asymptotic complexity
with a slightly larger constant factor.

The assembly of source terms and explicit evaluation of nonlinear coefficients
scale linearly with the number of degrees of freedom and therefore do not alter
the overall complexity.

\section{Setting up wildfire simulation}

The wildfire simulator has been setup using the IGA-ADS environment \cite{LOS201799}. 
wIn this section we discuss how to setup the wildfire simulation based on the model problem employed for the verification of the method. We assume a constant uniform distribution of the fuel and an initial circular ignition, which leads to a ring of fire propagating equally into all directions.

For the verification of the correctness of the code, we set up a model problem with zero wind, a ball-shaped initial temperature distribution in the center, 
and a constant uniform fuel distribution  equal to 1.
The model parameters are set up according to Table \ref{tab:parameters}. 

\begin{figure}[h!]
\centering
\begin{tabular}{cccc}
  \includegraphics[width=0.32\textwidth]{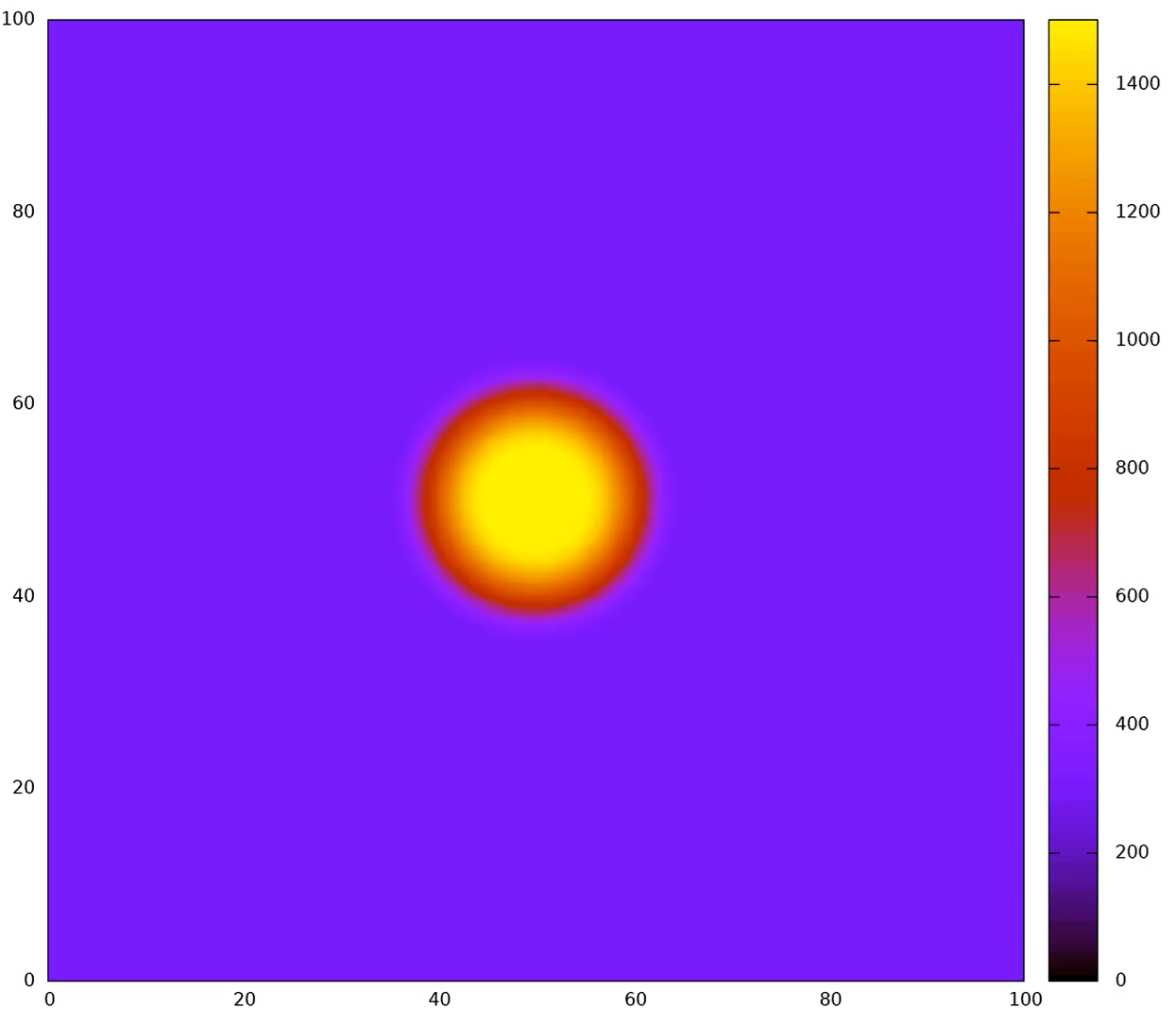} &
  \includegraphics[width=0.32\textwidth]{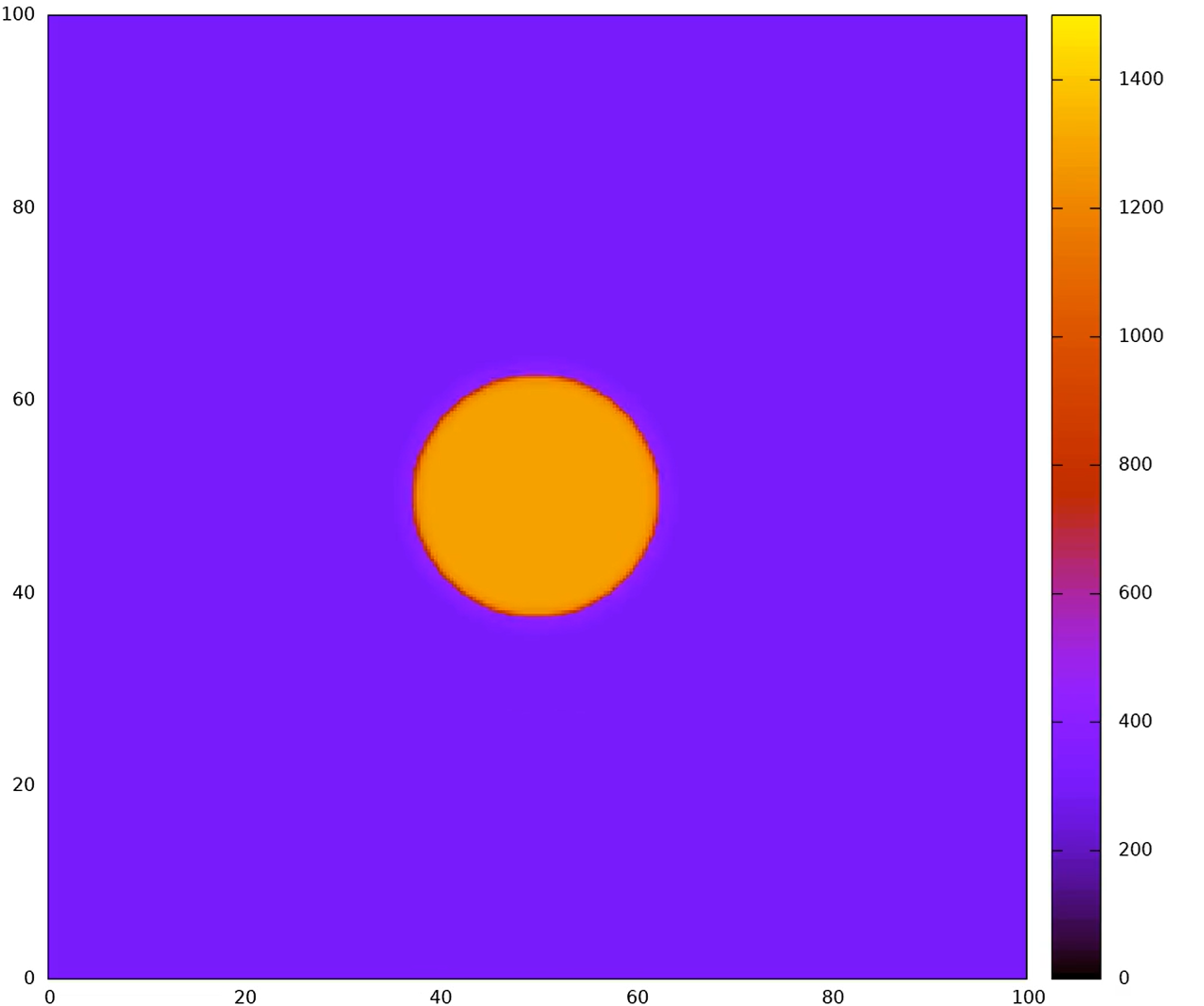} &
  \includegraphics[width=0.32\textwidth]{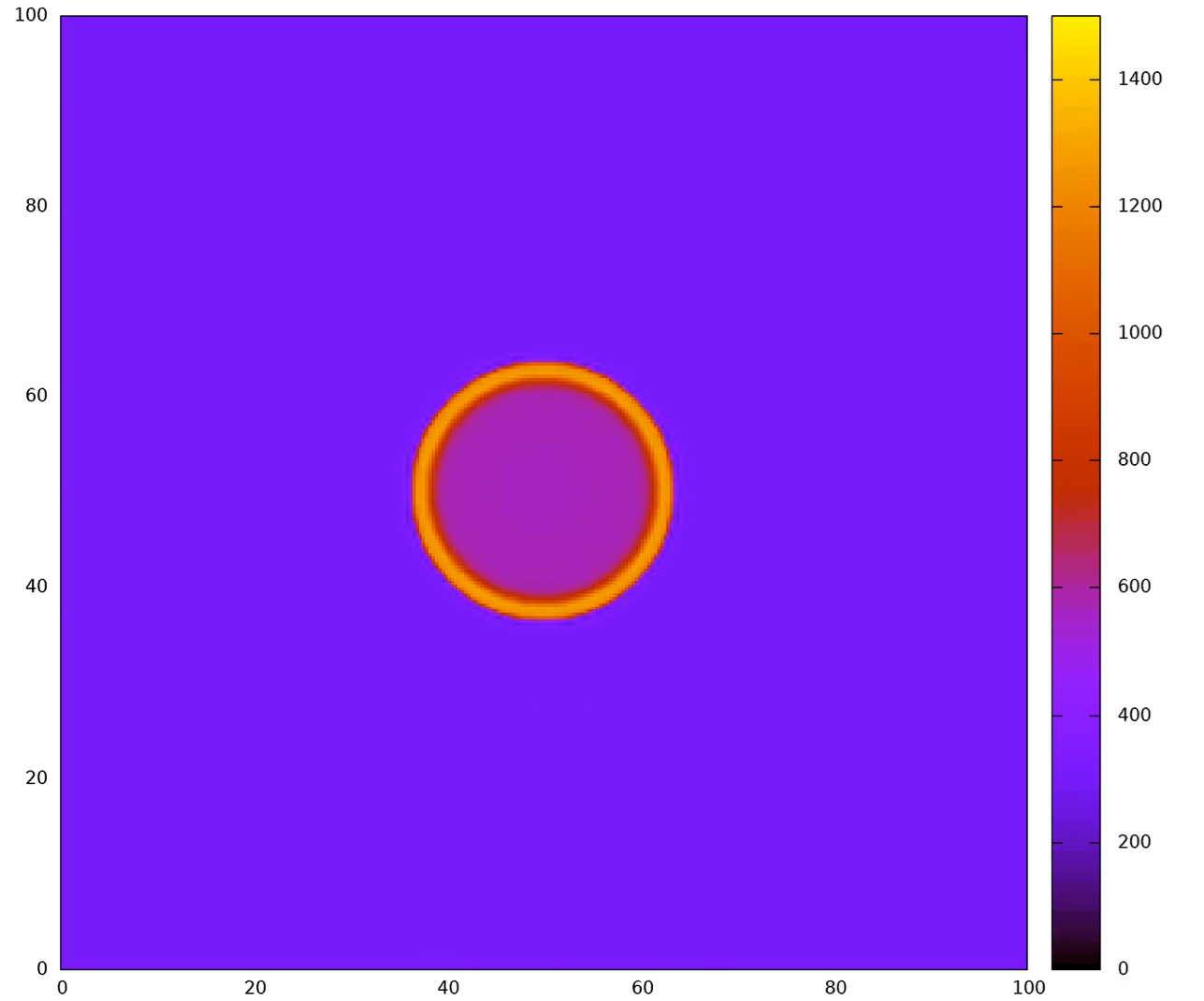} \\
  \includegraphics[width=0.32\textwidth]{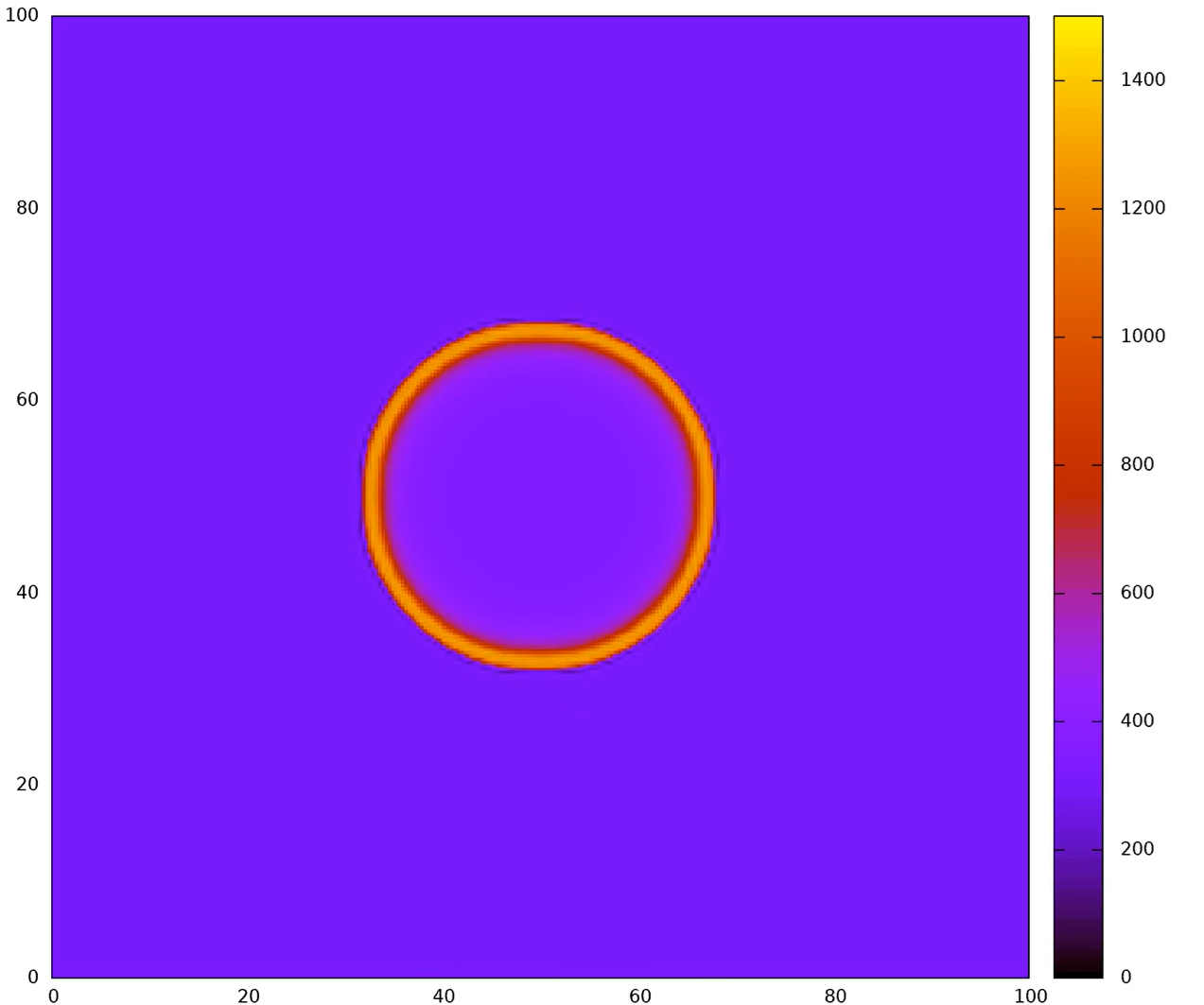} &
  \includegraphics[width=0.32\textwidth]{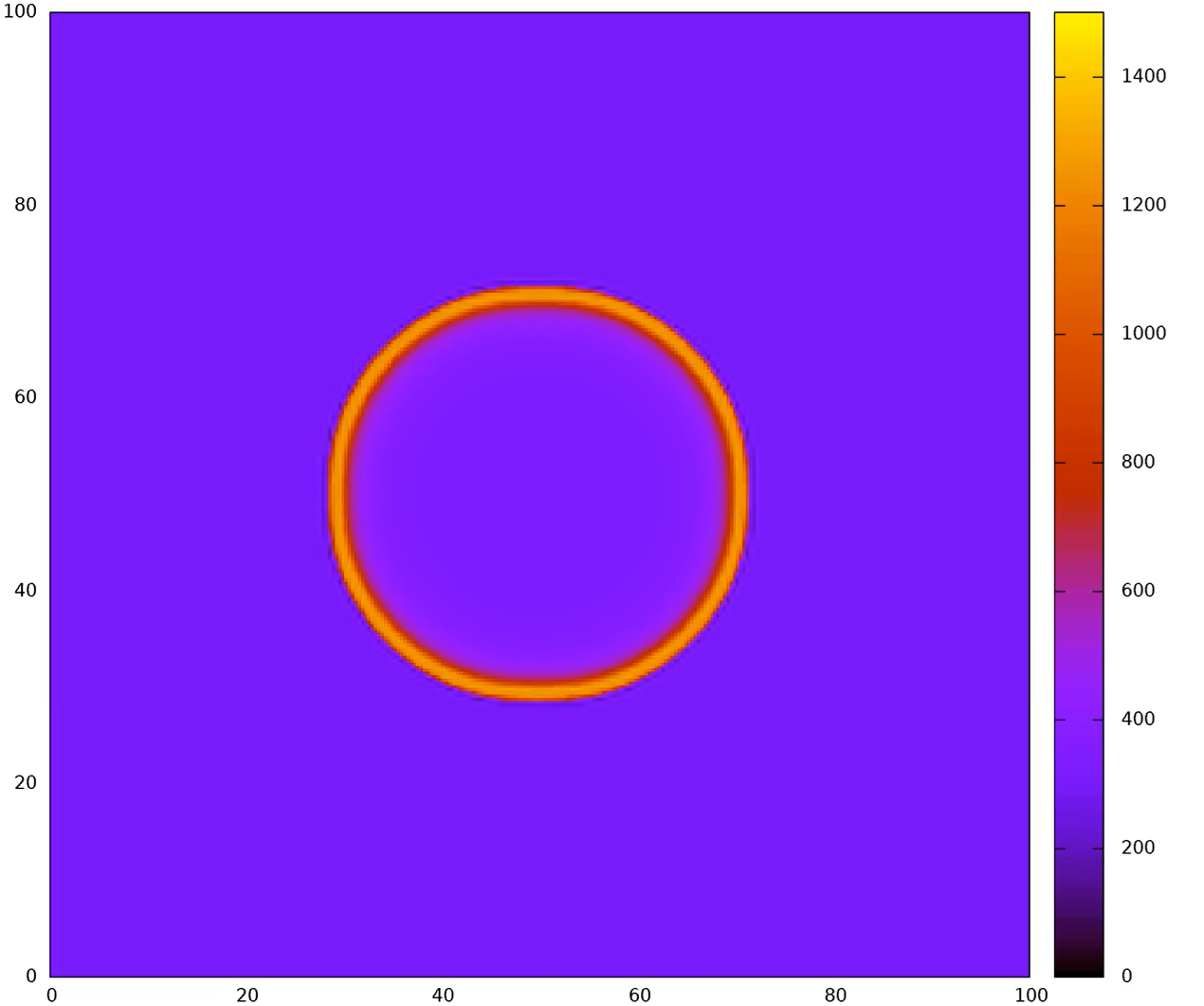} &
  \includegraphics[width=0.32\textwidth]{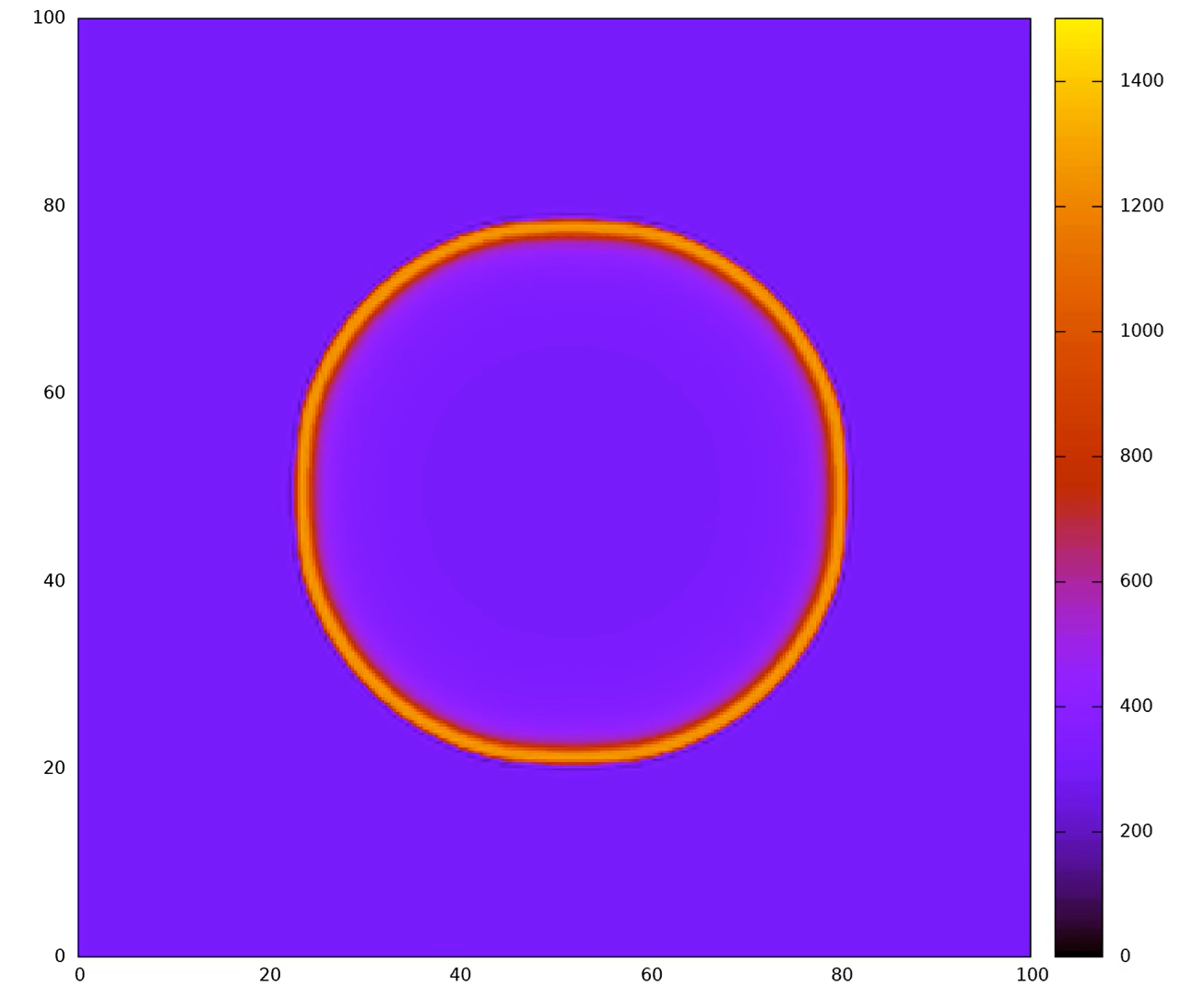} 
\end{tabular}
\caption{The spread of the wildfire as computed by the solver. }
\label{fig:verification}
\end{figure}

To set up a specific wildfire simulation, it is necessary to provide the following data
\begin{itemize}
\item the wind direction data in {\tt compute\_rhs} routine. The routine operates on the physical coordinates of the computational domain within the range $[0,100] \times [0,100]$. The physical point {\tt x[1], x[2]} is estimated by {\tt auto x = point(e,q)} for a specific element {\tt e} and the quadrature point {\tt q}. For this particular location, we provide the wind direction {\tt double bx(x), double by(x)}. For the simple circular simulation presented in this paper, we set up the zero wind.

\item the initial temperature configuration is defined in the {\tt initial\_state} routine.
There is a predefined routine {\tt bump} with smoothness parameters {\tt r} and {\tt R} that defines a smooth initial shape of the temperature distribution.

\begin{lstlisting}[
    language=C++,
    caption={Initial state and fuel map.},
    basicstyle=\ttfamily,
    keywordstyle=\color{blue},
    commentstyle=\color{green},
    stringstyle=\color{red},
    frame=single,
    numbers=left,
    numberstyle=\tiny\color{gray},
    stepnumber=1,
    numbersep=5pt,
    backgroundcolor=\color{lightgray!25}
]
// Computes a smooth transition function
inline double falloff
    (double r, double R, double t) {
 // Inside inner radius, full effect    
    if (t < r)
        return 1.0; 
    if (t > R)
 // Outside outer radius, no effect    
        return 0.0; 
    double h = (t - r) / (R - r);
    // Smooth falloff
    return std::pow((h - 1) * (h + 1), 2);  
}
inline double bump
    (double r, double R, double x, double y) {
    double dx = x - 50;  // Shift x-coordinate
    double dy = y - 50;  // Shift y-coordinate
    // Normalize distance
    double t = std::sqrt(dx * dx + dy * dy) / 100;  
    // Apply falloff
    return falloff(r / 200, R / 200, t);  
}
double init_state(double x, double y) {
    double r=10; double R=30;
    // it employs T0=300; Tcomb=1200;
    return T0+Tcomb*bump(r,R,x,y);
\end{lstlisting}
\item An initial fuel configuration needs to be defined. It can be prescribed in the {\tt initial\_state} routine. In this example, we set it up to a constant value {\tt auto fuel\_init = [] (double x, double y)\{return 1\}}. In a general case, the initial fuel configuration can be read from a file, as illustrated in the example below
\begin{lstlisting}[
    language=C++,
    caption={Initial state and fuel map.},
    basicstyle=\ttfamily,
    keywordstyle=\color{blue},
    commentstyle=\color{green},
    stringstyle=\color{red},
    frame=single,
    numbers=left,
    numberstyle=\tiny\color{gray},
    stepnumber=1,
    numbersep=5pt,
    backgroundcolor=\color{lightgray!25}
]
    const auto fuel_map {
      get_fuel_map()
    };
    const auto size_y {
      std::size(fuel_map)
    };
    const auto size_x {
      std::size(fuel_map[0])
    };
    const auto scale_x {
      static_cast < double > 
        (size_x) / (x.b - x.a)
    };
    const auto scale_y {
      static_cast < double > 
        (size_y) / (y.b - y.a)
    };
    auto fuel_init = [ & ](double x, double y) {
      const auto map_x {
        static_cast < std::size_t > (scale_x * x)
      };
      const auto map_y {
        static_cast < std::size_t > 
          (scale_y * (Base::y.b - y))
      };
      if (map_x >= size_x || map_y >= size_y)
        throw std::runtime_error
          (''Invalid map coordinates@'');
      return fuel_map[map_y][map_x] * 0.725;
    };
\end{lstlisting}

\item Finally, the solutions obtained for the fire and fuel terms are stored in {\tt out\_\%d.data} and {\tt fuel\_\%d.data} terms in the {\tt after\_step} routine.
\begin{lstlisting}[
    language=C++,
    caption={Code.},
    basicstyle=\ttfamily,
    keywordstyle=\color{blue},
    commentstyle=\color{green},
    stringstyle=\color{red},
    frame=single,
    numbers=left,
    numberstyle=\tiny\color{gray},
    stepnumber=1,
    numbersep=5pt,
    backgroundcolor=\color{lightgray!25}
]
// Perform tasks after each step
  void after_step(int iter, double /*t*/) override {
    auto i = iter  +1;  // Step index
    if (i % 10 == 0) {  // Output every 10 steps
        std::cout << ''Step '' << i << std::endl;
        output.to_file(u, ''out_%d.data'', i); 
        output.to_file(fuel, ''fuel_%d.data'', i); 
      }
\end{lstlisting}

\end{itemize}

For the model simulation presented in this section, we expect the wildfire to burn the fuel inside the fireball and to propagate as an ideal circle.
Numerical results presented in Figure \ref{fig:verification} confirm this behavior.

\subsection{Compiling and running the wildfire simulator}

The source code for the wildfire simulation is available on Github\footnote{https://github.com/marcinlos/iga-ads}.
To make setting up the code easy,
we provide a container definition and a set of recipes
for the task runner\footnote{https://github.com/casey/just}.
Using these, the entire installation process is simple
\begin{cmd}
    git clone https://github.com/marcinlos/iga-ads
    cd iga-ads
    just image
    just shell
\end{cmd}
and in the newly opened container shell:
\begin{cmd}
    just config
    just build
\end{cmd}
These commands build the simulation in the \texttt{/build} directory of the container.

The compiled simulation can be executed as
\begin{cmd}
    build/examples/fire <N> <p> <threads>
\end{cmd}
where \texttt{N} is the size of the mesh in each dimension,
\texttt{p} is the B-spline degree,
and \texttt{threads} is the number of threads to use.

\section{Numerical experiments}

\begin{figure}
    \centering
    \includegraphics[width=0.7\textwidth]{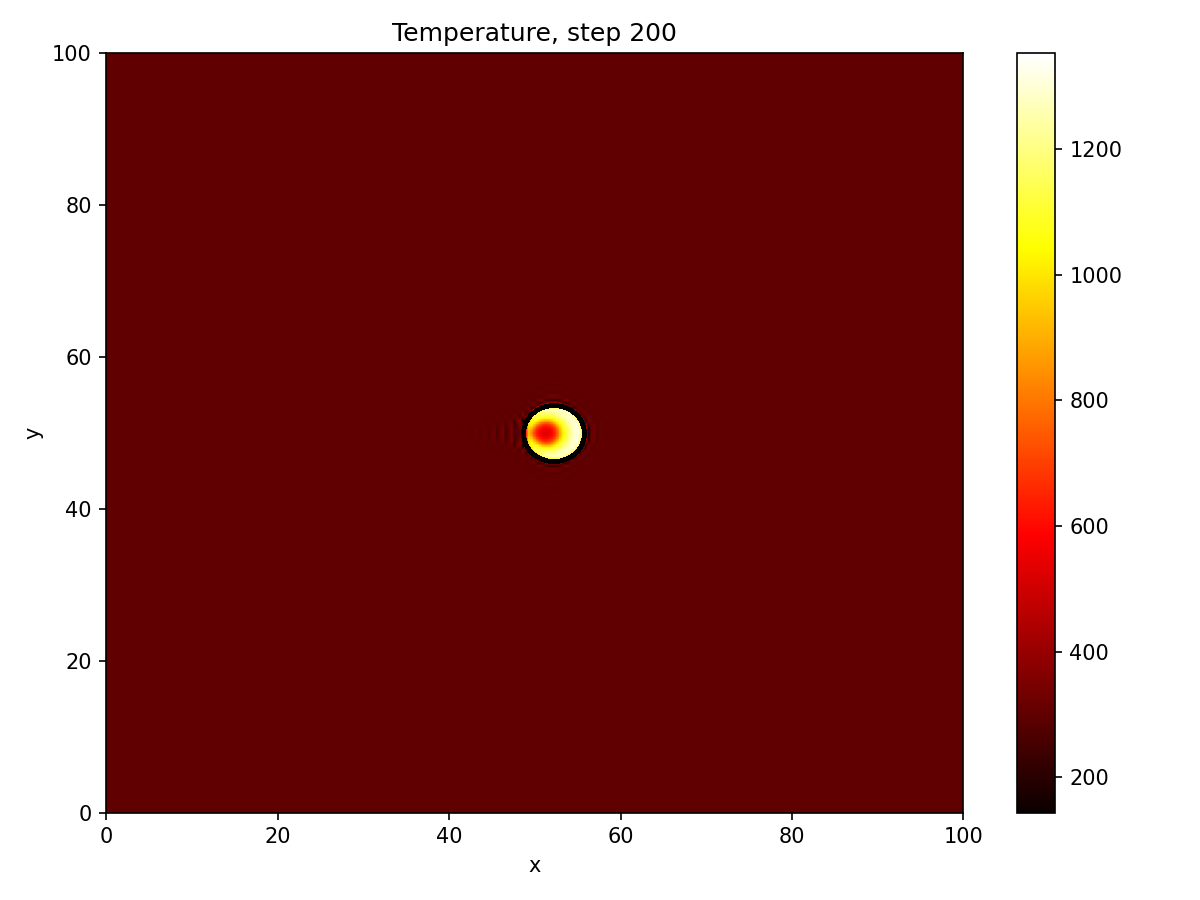}    \includegraphics[width=0.7\textwidth]{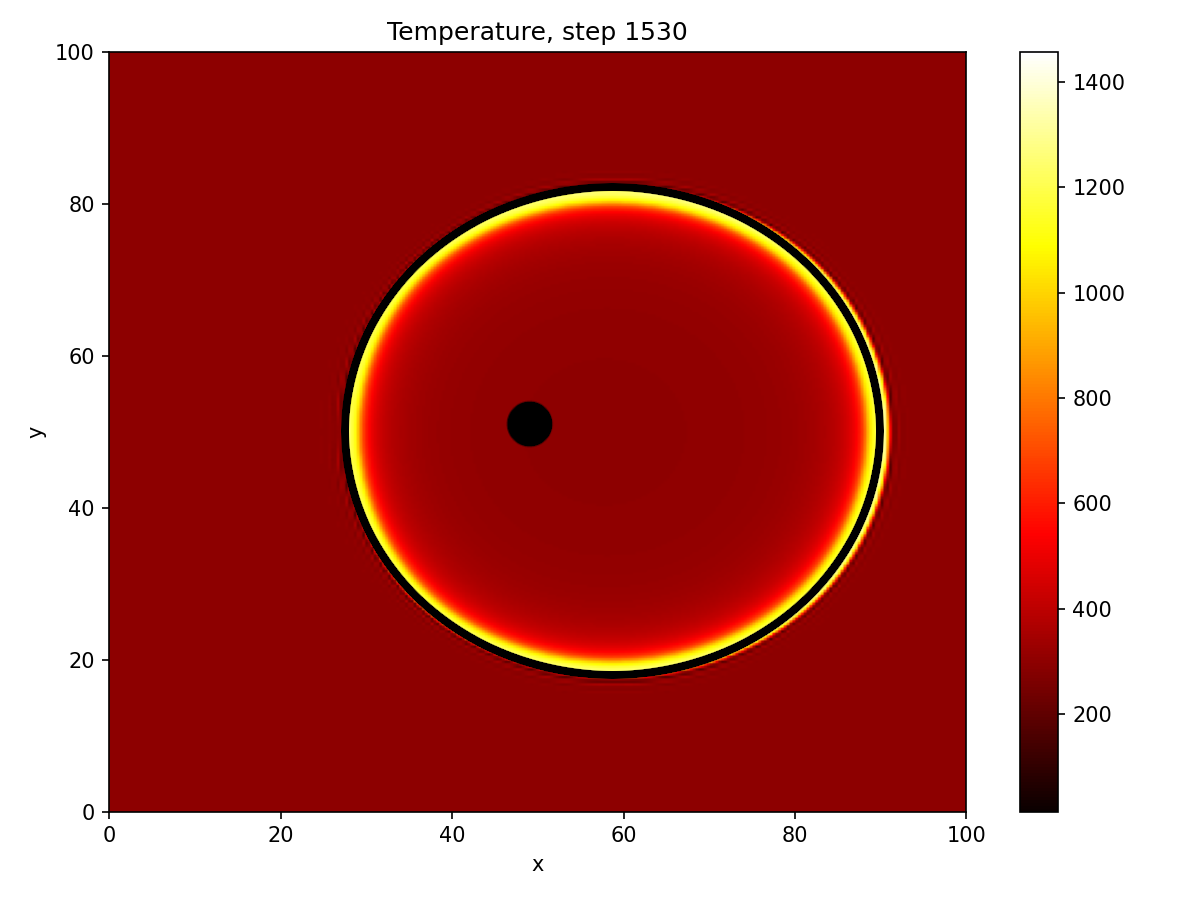}
    \caption{Comparison of the WILDFIRE-IGA-ADS simulations with the FAIRSITE simulations. The left panel denotes the initial state of the simulation. The right panel denotes the final state. The black dot there denotes the location of the initial ignition point. The black envelope denotes the final FAIRSITE fire front. The colors (from yellow, through red, to dark red) denotes the WILDFIRE-IGA-ADS final temperature distribution.}
    \label{fig:jajko}
\end{figure}

\revision{We start with the comparison of our code with the FARSITE wildfire simulator \cite{finney1998farsite}, see Figure \ref{fig:jajko}. The FAIRSITE front has been coppied from Figure 4 of \cite{w8}. We compare the simulations with a single initial ignition point, a uniform fuel map, and the wind in the right direction. We can observe a similar evolution of the fire front.}

\revision{In this section, we also present three numerical examples of the wildfire simulation.}
\revision{The first one is the verification of the stability and order of the time integration scheme for the model problem with manufactured solution, solved using Peaceman-Rachford and Strang with Crank-Nicolson time integration schemes. We test the accuracy and convergence of the method for linear and nonlinear problem setups.}

\begin{figure}[h]
    \centering
    \includegraphics[width=0.49\textwidth]{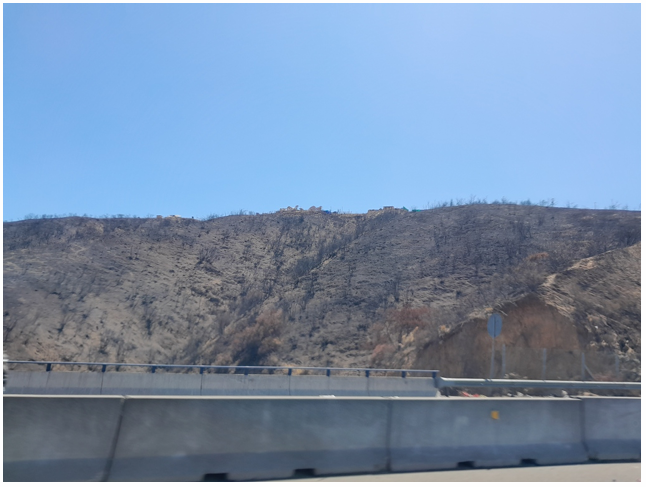} \includegraphics[width=0.47\textwidth]{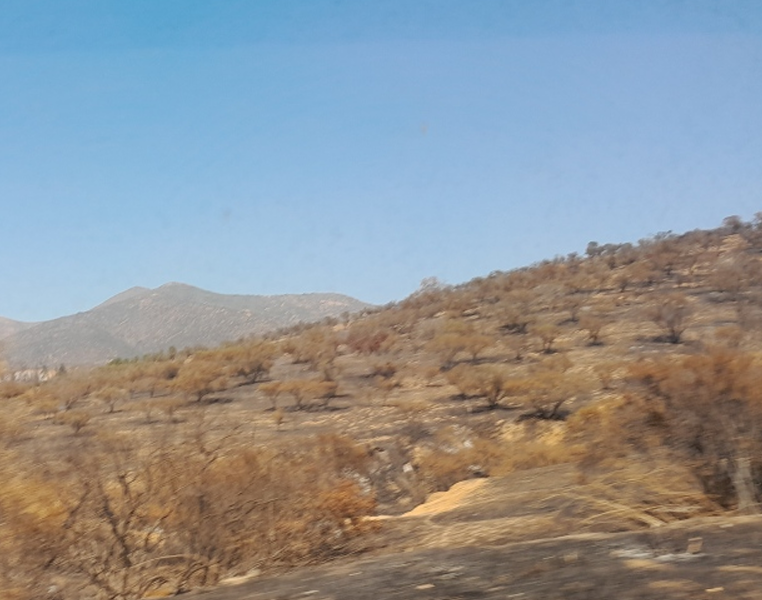}
    \caption{Burnt forest near the Valparaiso region of Chile. Photos made February 10, 2024 by Maciej Paszy\'nski from the highway affected by the wildfire.}
    \label{fig:photo}
\end{figure}

\begin{figure}[h]
    \centering
    \includegraphics[width=0.8\textwidth]{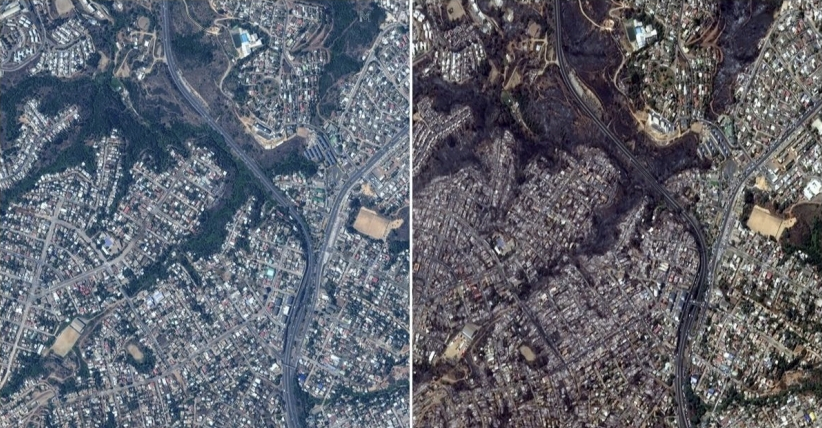}
    \caption{Comparison of a satellite image taken before and after the fire (from  \cite{cnnchilefires}).}
    \label{fig:chile_real}
\end{figure}

The second example concerns the wildfire that happened in the Valparaiso region of Chile. We simulate the situation that takes place at the beginning of 2024 in central and southern Chile. From February 1 to 5, 2024, a series of wildfires occurred. Figure \ref{fig:photo} presents the photo of the burned forest taken from the highway in the Valparaiso region of Chile by Maciej Paszy\'nski, a week after the wildfire ended. According to a Chilean government statement, this incident is the worst disaster since the 2010 earthquake. There were 162 forest fires, and several regions of Chile were affected, including Valparaiso, O’Higgins, Maule, Biobío, and Los Lagos. The causes of these disasters were high temperatures worsened by the El Niño phenomenon. In total, 43,000 hectares were affected by the wildfires, and 130 people were killed \cite{cnnchilefires}. For the sake of simulation, we chose a specific urban region, Viña del Mar. The satellite images of this region from \cite{cnnchilefires} are presented in Figure \ref{fig:chile_real}.

The third example concerns the wildfire that happened on the island of Las Palmas in Gran Canaria between August 17 and 20, 2019. The phenomenon is well-reported. There are European Space Agency reports \cite{ESACanaries}, including satellite photos from the IR channels, for August 17, 18, and 19 \cite{Canaries1}, as well as for August 20 \cite{Canaries2}. The wildfire lasted 5 days and burned 25,000 acres, including parts of Tamadaba National Park.

\subsection{\revision{Stability and order of the time integration scheme}}

\revision{In this section, we performed the following numerical experiments.
We select the mesh dimensions as $50\times 50$ and $100 \times 100$ using quadratic B-splines.
The computational domain is $[0,100]^2$.
We change the time step size from 1, 1/2, 1/4, 1/8, 1/16, 1/32, 1/64, 1/128.
We run the number of times such that it covers the time interval [0,1].
We solve the following equations. First, we solve the explicit dynamics problem
\begin{equation}
\begin{aligned}
\underbrace{ \frac{\partial T}{\partial t}}_{\textrm{\textrm{time progression}}} = & -{\underbrace{C_{\textrm{advection}}  \mathbf{b} \cdot \nabla T}_{\textrm{advection}} +\underbrace{C_{\textrm{diffusion}} \nabla \cdot  \nabla T}_{\textrm{diffusion}} +\underbrace{C_{\textrm{reaction}}   T}_{\textrm{reaction}} } +f + \notag \\
&\ingray{\underbrace{\nabla \cdot (C_{\textrm{non-linear diffusion}} \cdot T^3 \nabla T)}_{\textrm{non-linear diffusion}}} +
\notag \\ & \ingray{ C_{\textrm{ignition}} \cdot [T>T_{ig} \textrm{ AND fuel}>0.2]T \exp(\frac{-300}{T})+}  
\\  & \ingray{\underbrace{ \qquad \qquad \qquad \qquad \qquad + C_{\textrm{forcing}}- C_{\textrm{radiation}} \cdot  T^4}_{\textrm{non-linear forcing}}}.
\end{aligned} \label{eq:linear}
\end{equation}
Second, we solve the non-linear problem (with the non-linear terms on the right-hand-side).
\begin{equation}
\begin{aligned}
\underbrace{ \frac{\partial T}{\partial t}}_{\textrm{\textrm{time progression}}} +& {\underbrace{C_{\textrm{advection}}  \mathbf{b} \cdot \nabla T}_{\textrm{advection}} -\underbrace{C_{\textrm{diffusion}} \nabla \cdot  \nabla T}_{\textrm{diffusion}} -\underbrace{C_{\textrm{reaction}}   T}_{\textrm{reaction}} } = f +\\
&\ingray{\underbrace{\nabla \cdot (C_{\textrm{non-linear diffusion}} \cdot T^3 \nabla T)}_{\textrm{non-linear diffusion}}} +
\notag \\ & \ingray{ C_{\textrm{ignition}} \cdot [T>T_{ig} \textrm{ AND fuel}>0.2]T \exp(\frac{-300}{T})+}  
\\  & \ingray{\underbrace{ \qquad \qquad \qquad \qquad \qquad + C_{\textrm{forcing}}- C_{\textrm{radiation}} \cdot  T^4}_{\textrm{non-linear forcing}}}.
\end{aligned} \label{eq:nonlinear}
\end{equation}
For each problem we select a manufactured solution of the form 
\begin{equation} u_{exact}(x,y;t)=300+80 \hat{u}_{1,1,3}+30 \hat{u}_{2,1,5}+ 110 \hat{u}_{2,2,1.5},
\end{equation}
 where
\begin{equation}
\hat{u}_{n_x,n_y,\lambda}(x,y;t)=\left(1-\cos(2 \Pi \frac{x}{100})\right)\left(1-\cos(2 \Pi \frac{y}{100})\right)\exp(-\lambda t),
\end{equation}
with zero-Neuman b.c. We enforce the manufactured solution by computing the forcing term $f$. As the initial configuration, we setup
\begin{equation}
u(x,y;0)=u_{exact}(x,y,0).
\end{equation}
For problem \eqref{eq:linear} we employ the explicit dynamics solver with the Backward Euler method.
For problem \eqref{eq:nonlinear} we employ the Peaceman-Rachford time integration scheme, and the Strang splitting scheme with the Crank-Nicolson method.
Summing up, we perform the following sequence of experiments:
\begin{enumerate}
\item Backward-Euler explicit time integration scheme for linear equations \eqref{eq:linear}, for $50\times 50$, $100\times 100$, and $200 \times 200$ meshes with quadratic B-splines, solved for the time step sizes 1, 1/2, 1/4, 1/8, 1/16, 1/32, 1/64, 1/128 partitioning the [0,1] interval.
\item Peaceman-Rachford time integration scheme for non-linear equations \eqref{eq:nonlinear}, for $50\times 50$, $100\times 100$, and $200 \times 200$ meshes with quadratic B-splines, solved for the time step sizes 1, 1/2, 1/4, 1/8, 1/16, 1/32, 1/64, 1/128 partitioning the [0,1] interval.
\item Strang splitting scheme with Crank-Nicolson method for non-linear equations \eqref{eq:nonlinear}, for $50\times 50$, $100\times 100$, and $200 \times 200$ meshes with quadratic B-splines, solved for the time step sizes 1, 1/2, 1/4, 1/8, 1/16, 1/32, 1/64, 1/128 partitioning the [0,1] interval.
\end{enumerate}}

\begin{figure}
\centering
  \includegraphics[width=0.9\textwidth]{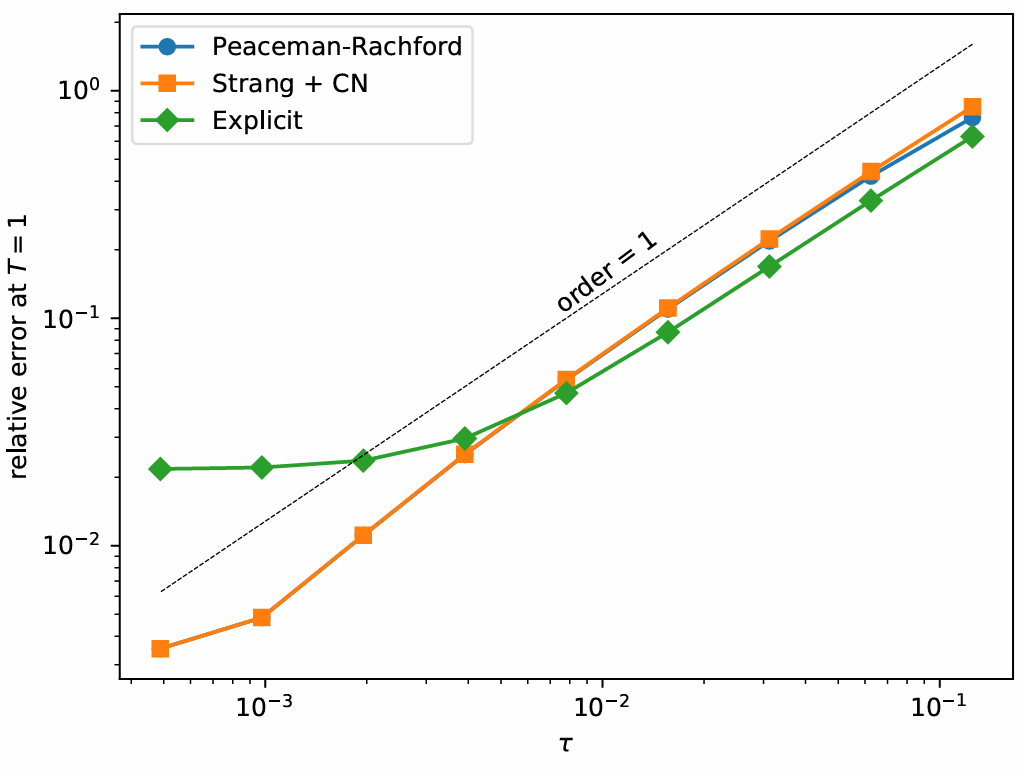} 
  \caption{Comparison of convergence of the explicit scheme and the quasi-implicit schemes, for Peaceman-Rachford and Strang with Crank-Nicolson time integration scheme. Quadratic B-splines of $C^1$ continuity, $50 \times 50$ mesh.}
  \label{fig:convergence1}
\end{figure}

\begin{figure}[h!]
\centering
  \includegraphics[width=0.9\textwidth]{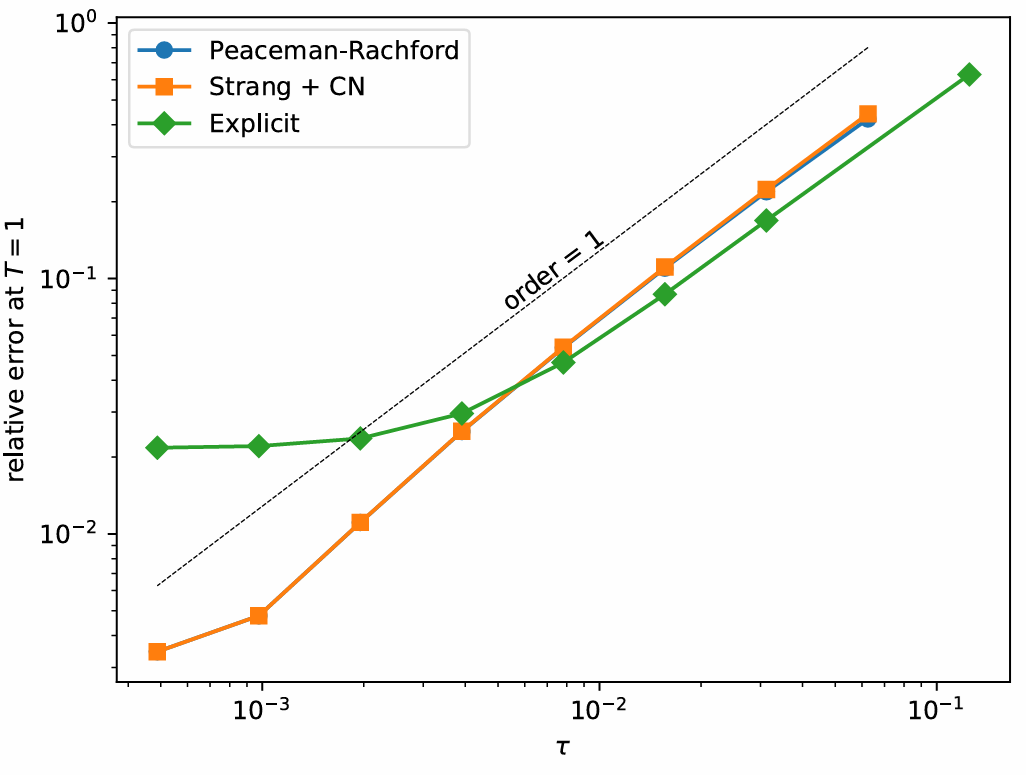} 
  \caption{Comparison of convergence of the explicit scheme and the quasi-implicit schemes, for Peaceman-Rachford and Strang with Crank-Nicolson time integration scheme. Quadratic B-splines of $C^1$ continuity, $100 \times 100$ mesh.}
  \label{fig:convergence2}
\end{figure}

\begin{figure}[h!]
\centering
  \includegraphics[width=0.9\textwidth]{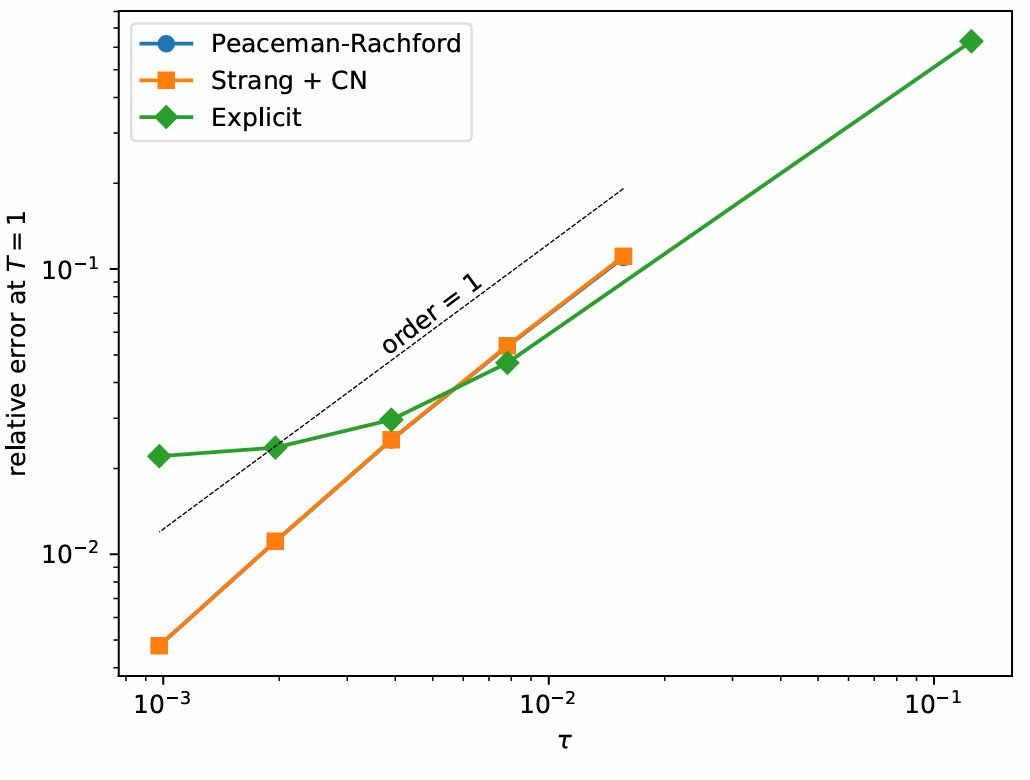} 
  \caption{Comparison of convergence of the explicit scheme and the quasi-implicit schemes, for Peaceman-Rachford and Strang with Crank-Nicolson time integration scheme. Quadratic B-splines of $C^1$ continuity, $200 \times 200$ mesh.}
  \label{fig:convergence3}
\end{figure}

\revision{The measurements are presented in Figures \ref{fig:convergence1}-\ref{fig:convergence3}. The horizontal axis denotes the time step size, the vertical axis denotes the relative error computed over the entire [0,1] time interval, between the numerical result and the known exact solution. The points that are absent indicate an unstable solution (large error). We can draw the following conclusions from these measurements. All the schemes are of the first order. Unfortunately, the schemes are not unconditionally stable. The quasi-implicit time integration scheme provides one order of numerical accuracy higher than explicit schemes for small time steps. There is a single artifact for a $200\times 200$ mesh, where the explicit scheme is stable for a time step of 1/2 (two time steps executed).}

\subsection{Wildfire at the Valparaiso region of Chile}


  Our wildfire solver can simulate the spread of fire on the given fuel map. The fuel map is a 2D array of double-precision numbers in the range $[0, 1]$. That input is created using the actual satellite image \ref{fig:chile_real}. Each pixel represents the amount of fuel in that area. The fuel map is presented in Figure \ref{fig:buttle}.
  As the base for our problem, we used two satellite images. To prepare the fuel map, we extracted the green pixels from the original image. We used a threshold for the green value to better approximate the lighting elements in the terrain.
  Then the values were converted to numbers in the range $[0.0, 1.0]$ relative to the intensity of the green color at a given point. This allowed us to use the processed image as a 2D data array in the next steps.
  With these assumptions, we have run the simulations.
  Figure \ref{fig:image_grid} presents snapshots from the simulation, starting from the initial configuration.
  As we can see, by comparing the wildfire paths on the presented snapshots in Figure \ref{fig:image_grid} with the post wildfire image presented in Figure \ref{fig:chile_real}, our method predicted the spread of the fire in the Viña del Mar arena; however, even the initial circular state assumed in our simulation was actually larger than the realistic scenario.

\begin{figure}[h!]
\centering
  \includegraphics[width=0.5\textwidth]{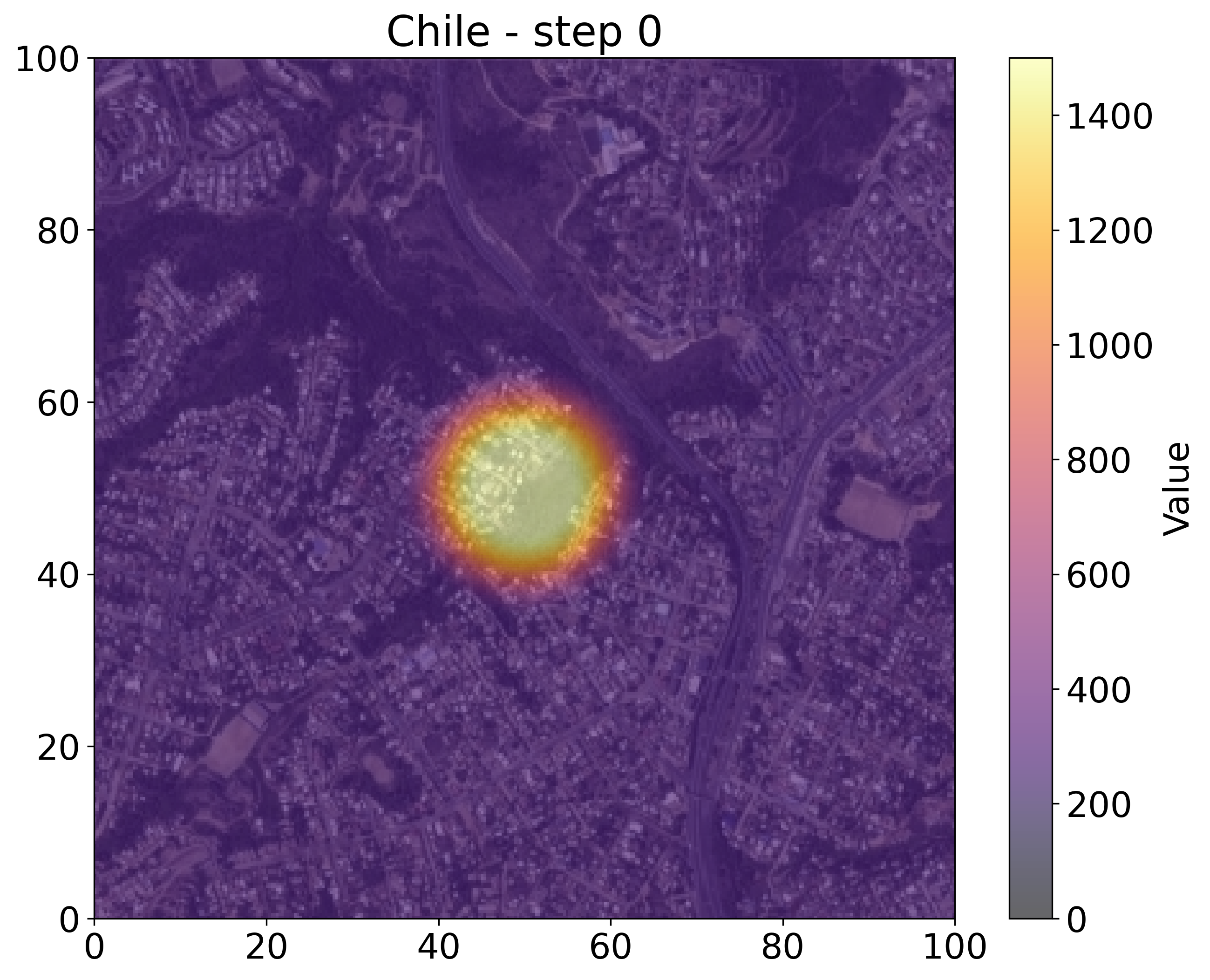} 
  \caption{Initial fire configuration for the Viña del Mar simulation.}
  \label{fig:initialvina}
\end{figure}

\begin{figure}[h!]
\centering
  \includegraphics[width=0.32\textwidth]{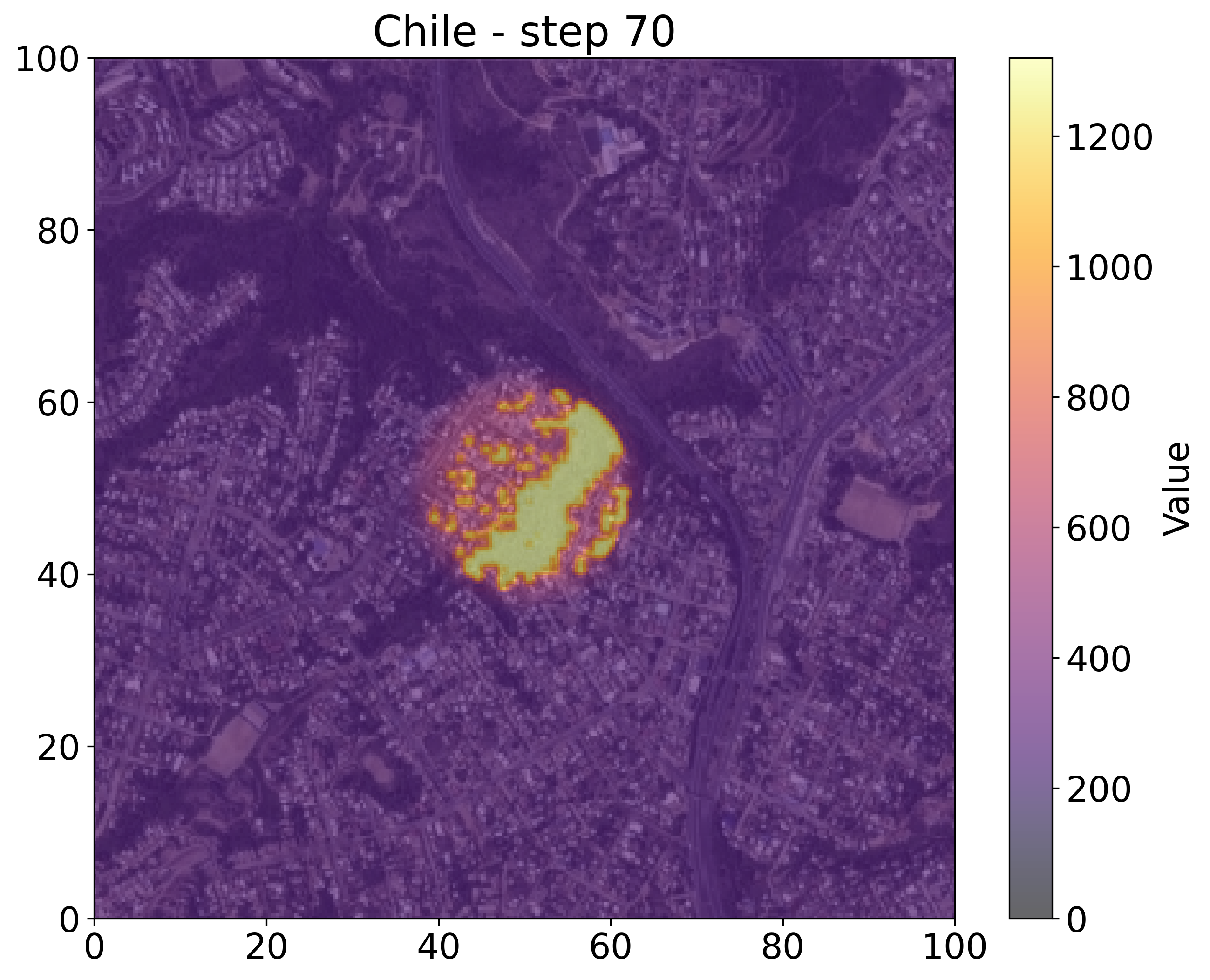}   \includegraphics[width=0.32\textwidth]{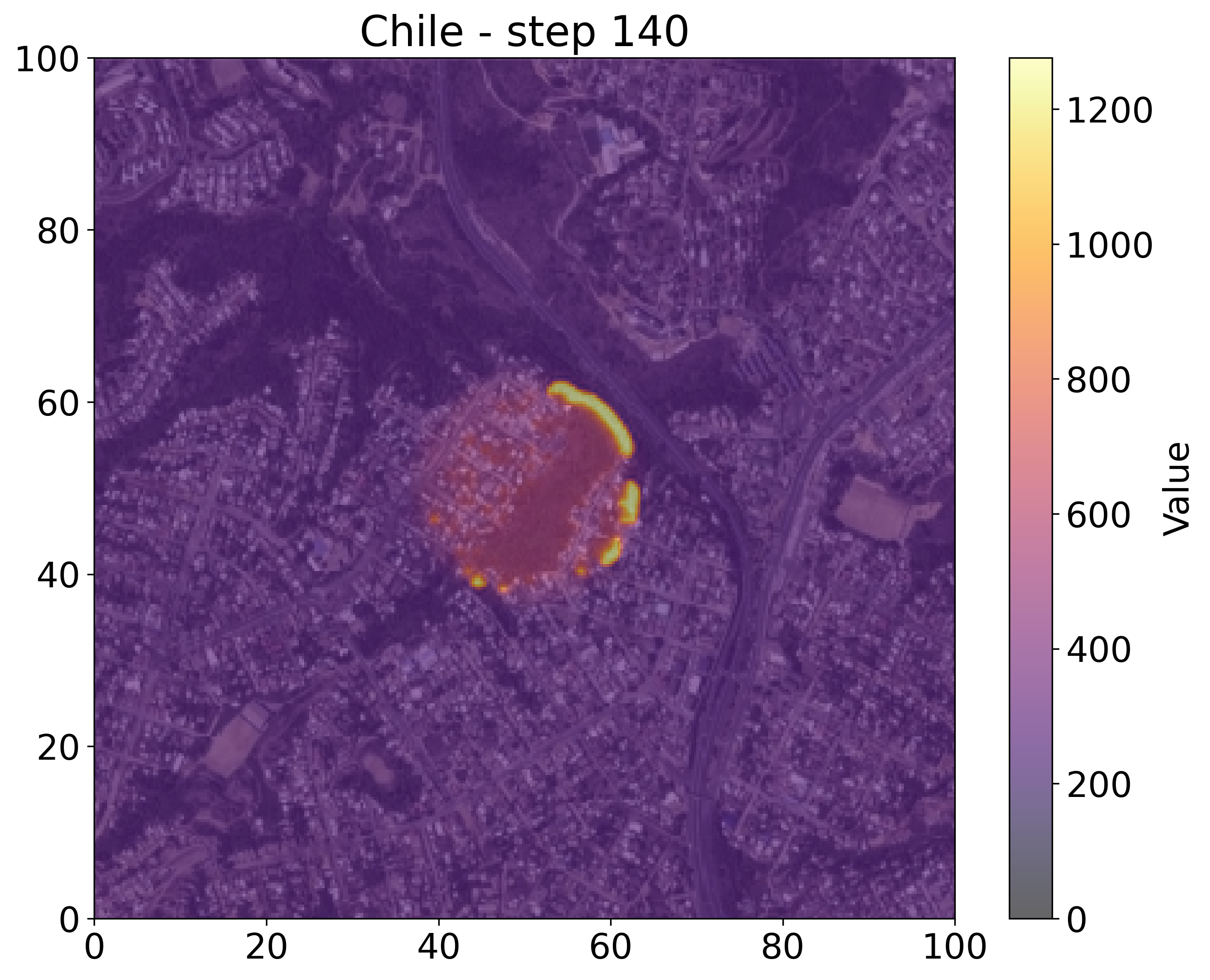} \includegraphics[width=0.32\textwidth]{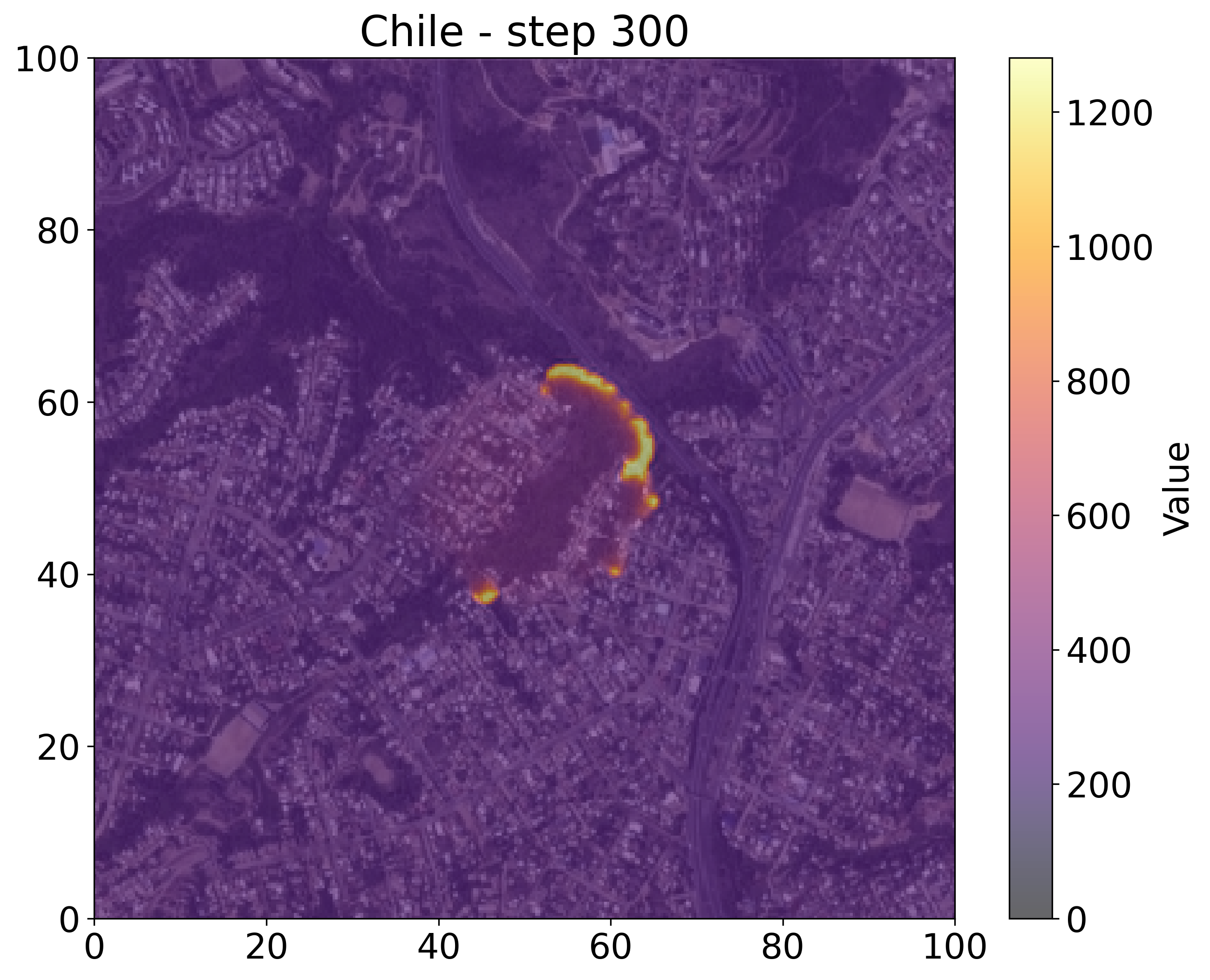}  \\
\includegraphics[width=0.32\textwidth]{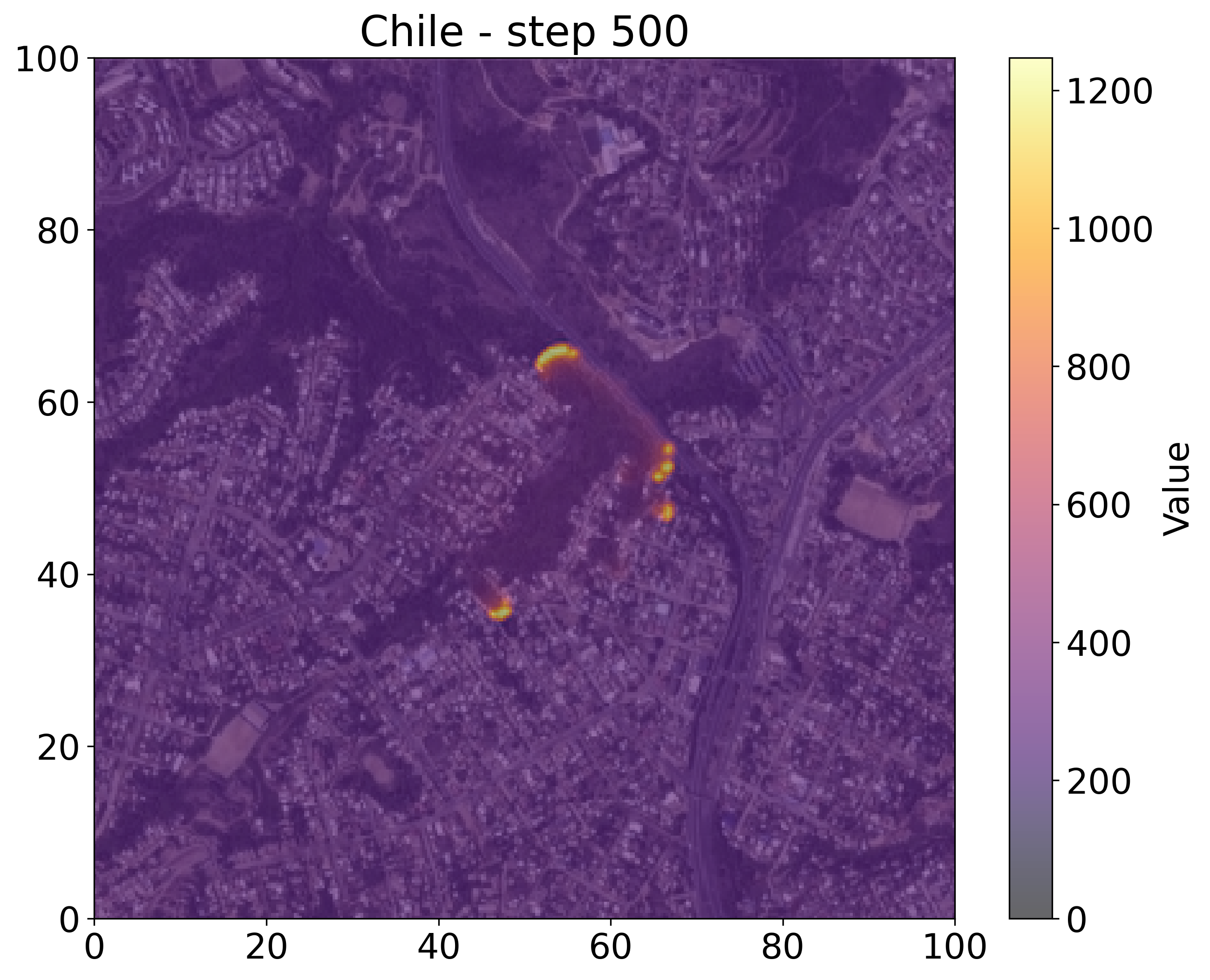}   \includegraphics[width=0.32\textwidth]{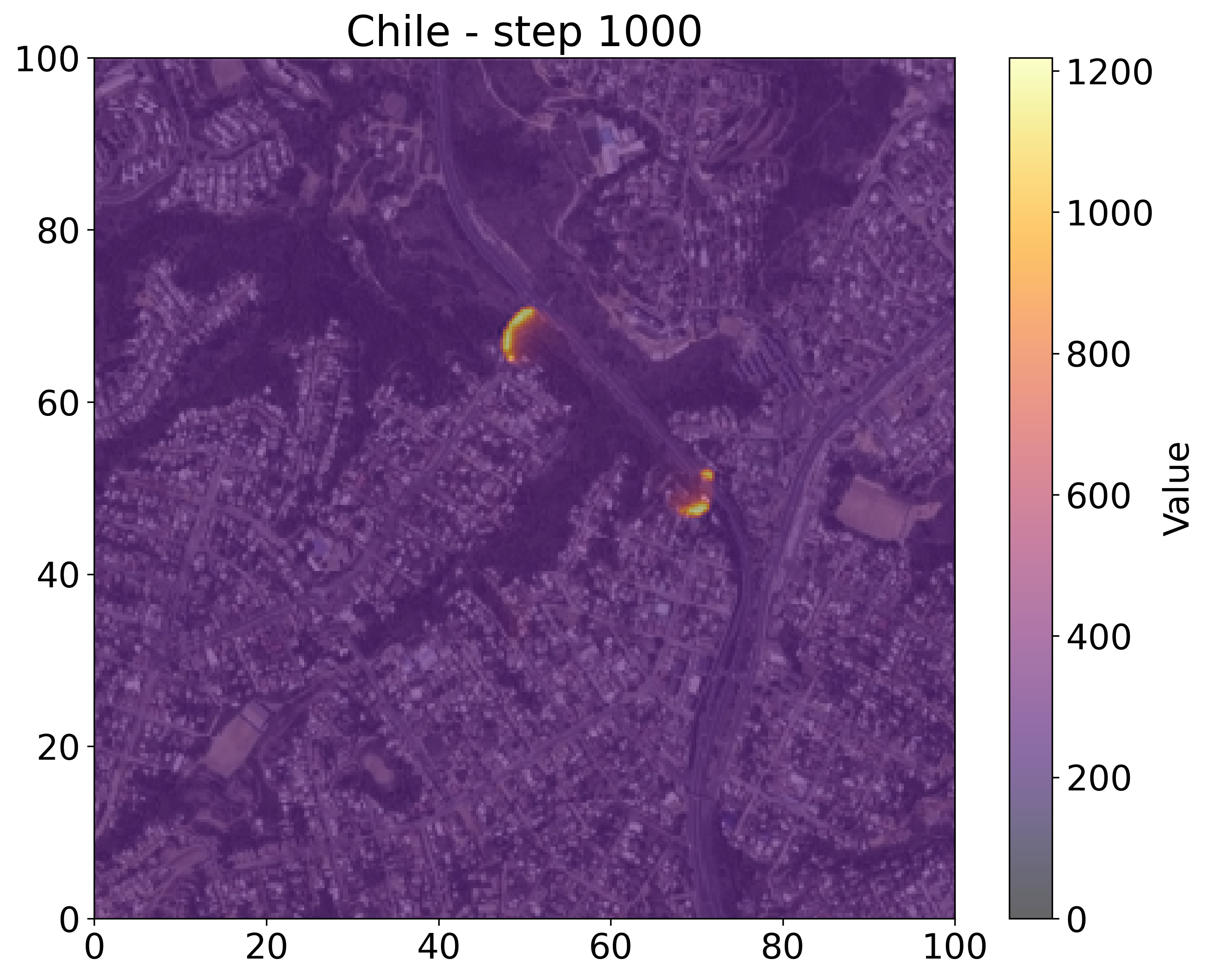} \includegraphics[width=0.32\textwidth]{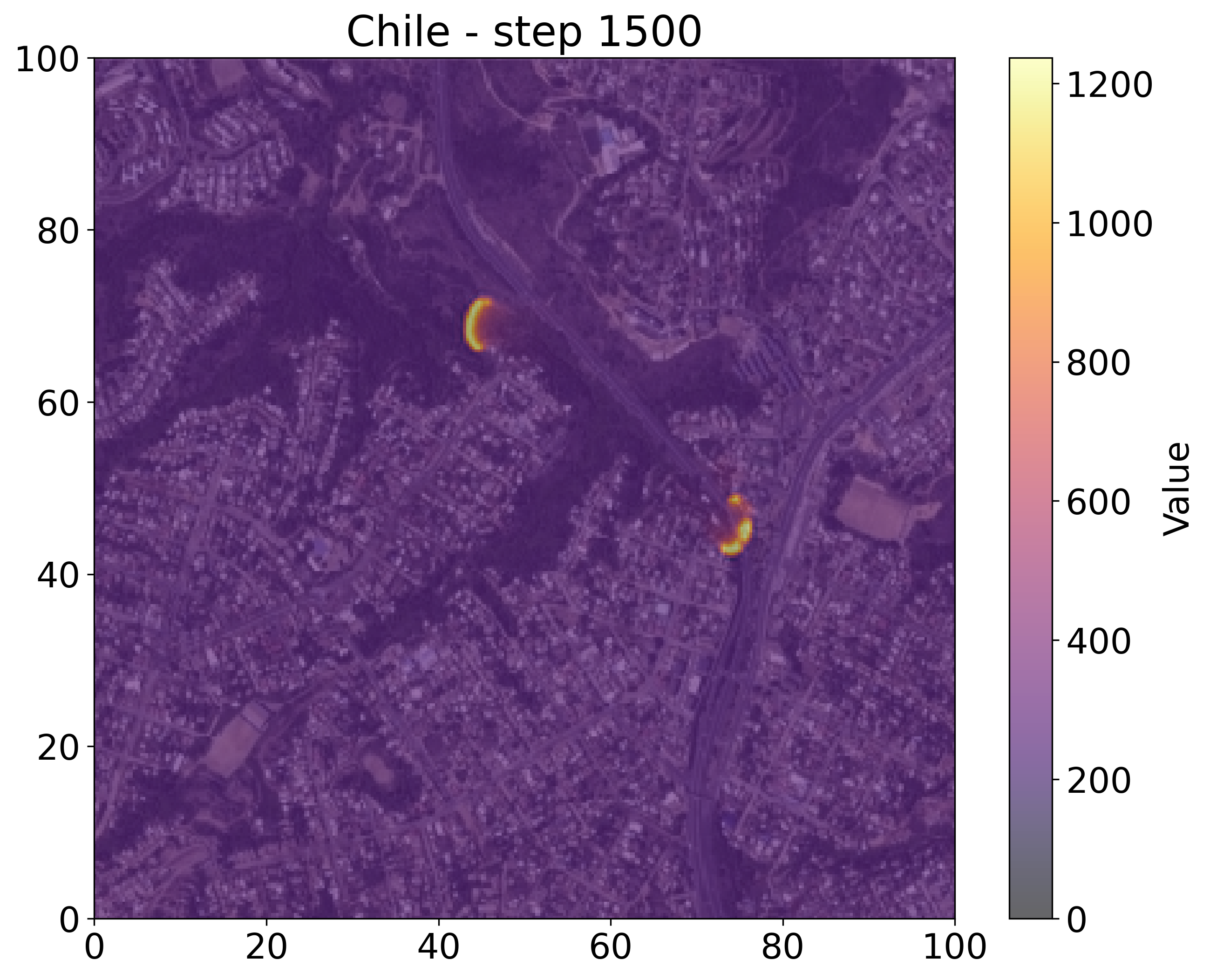}  \\
\includegraphics[width=0.32\textwidth]{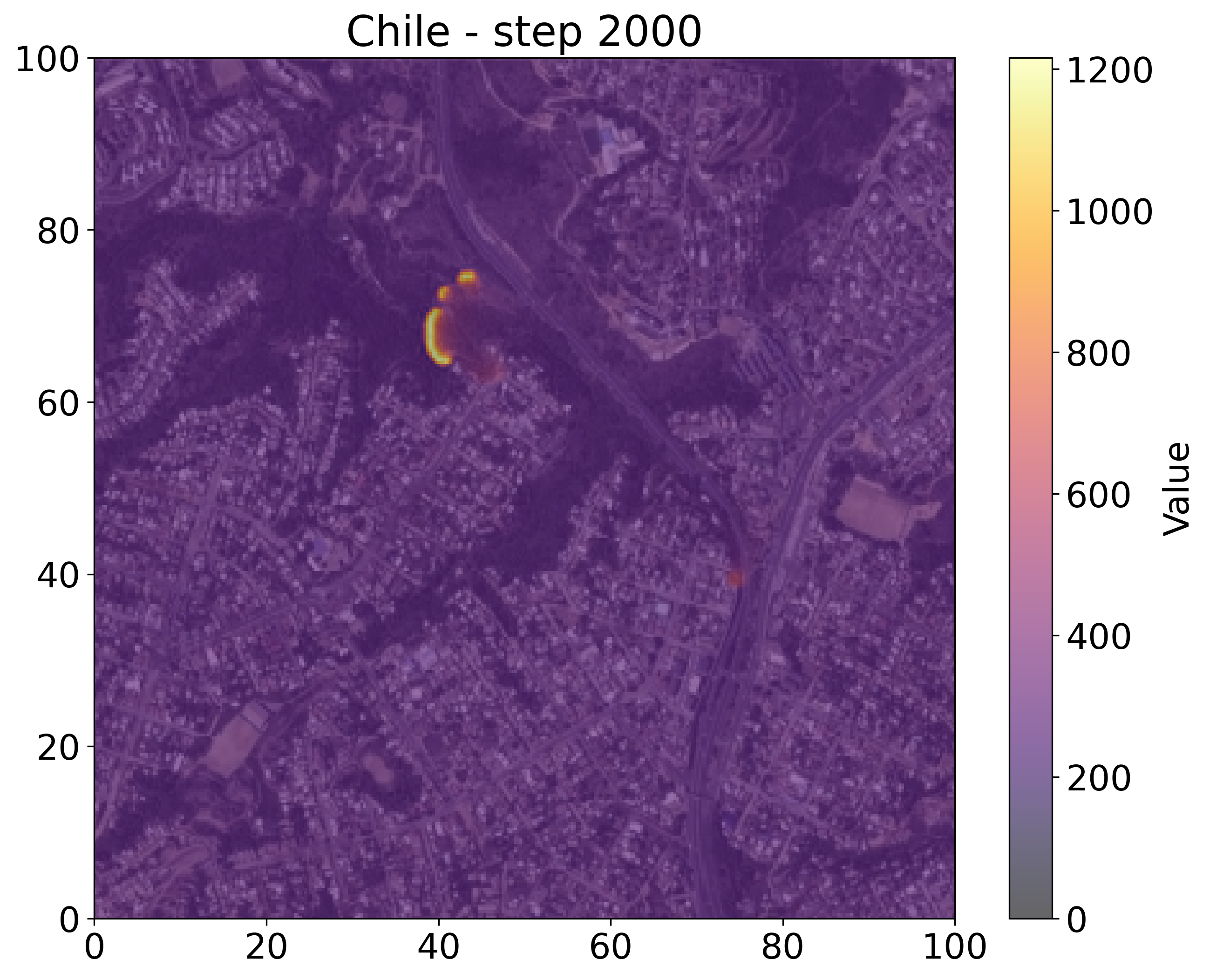}   \includegraphics[width=0.32\textwidth]{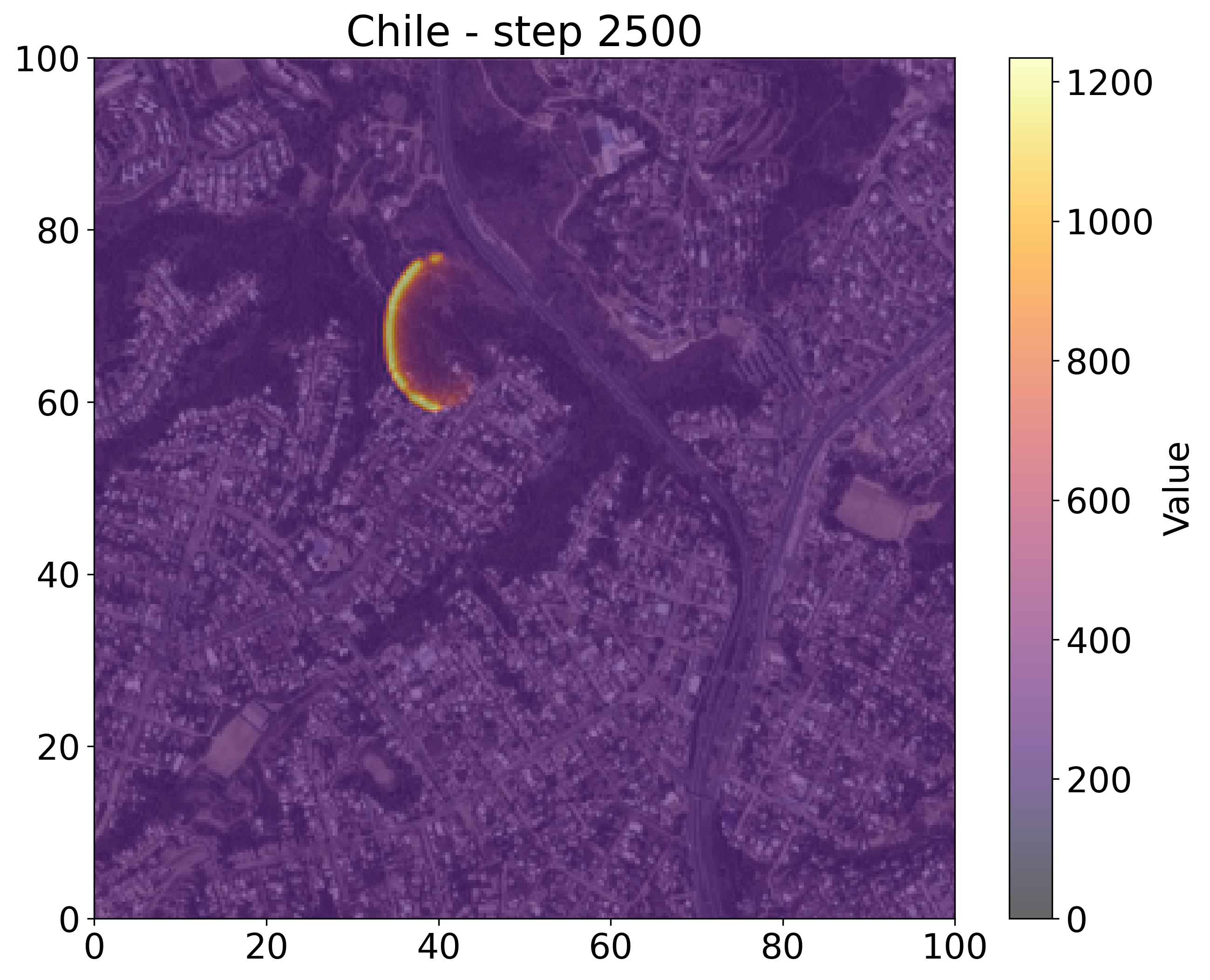} \includegraphics[width=0.32\textwidth]{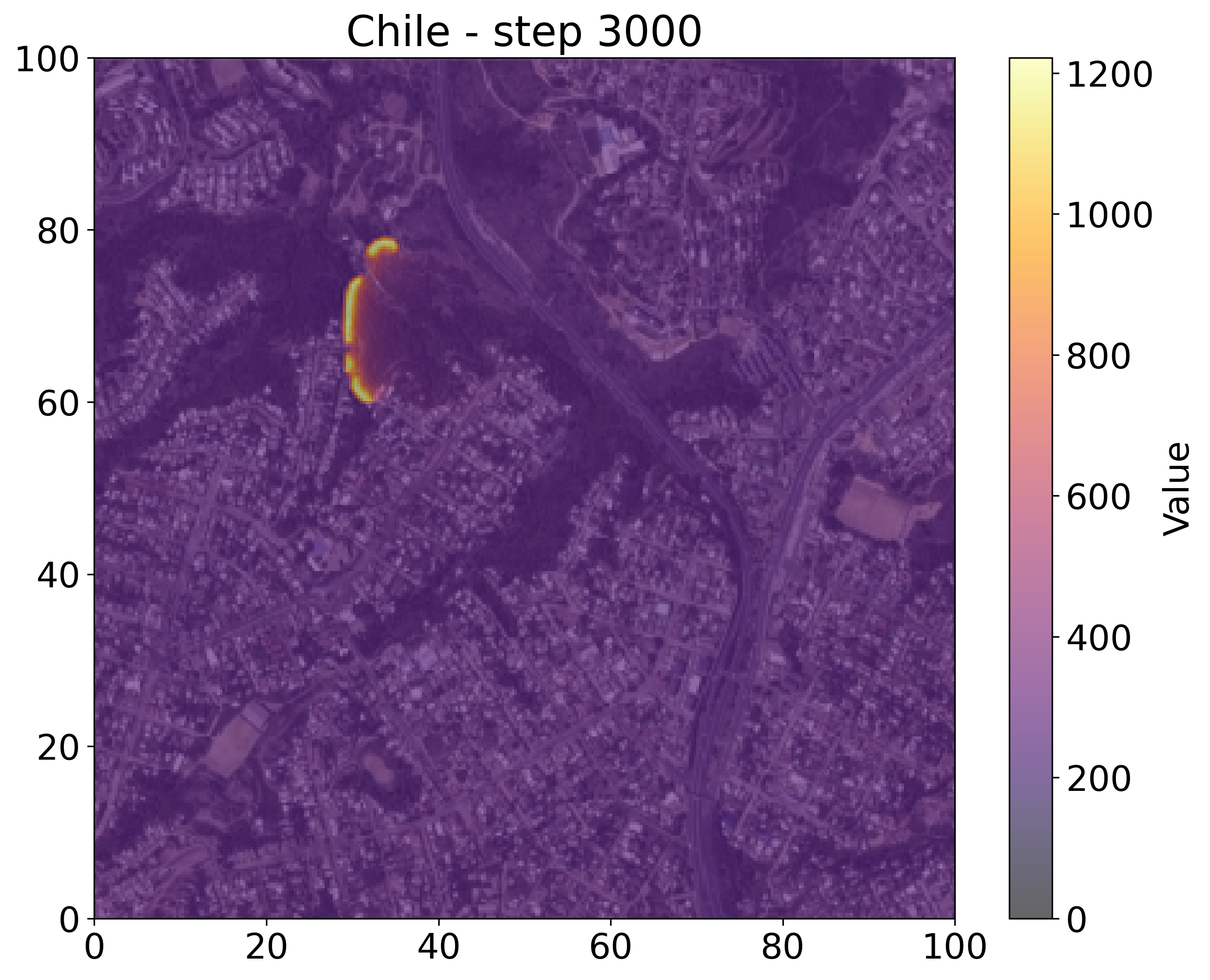}  \\
\includegraphics[width=0.32\textwidth]{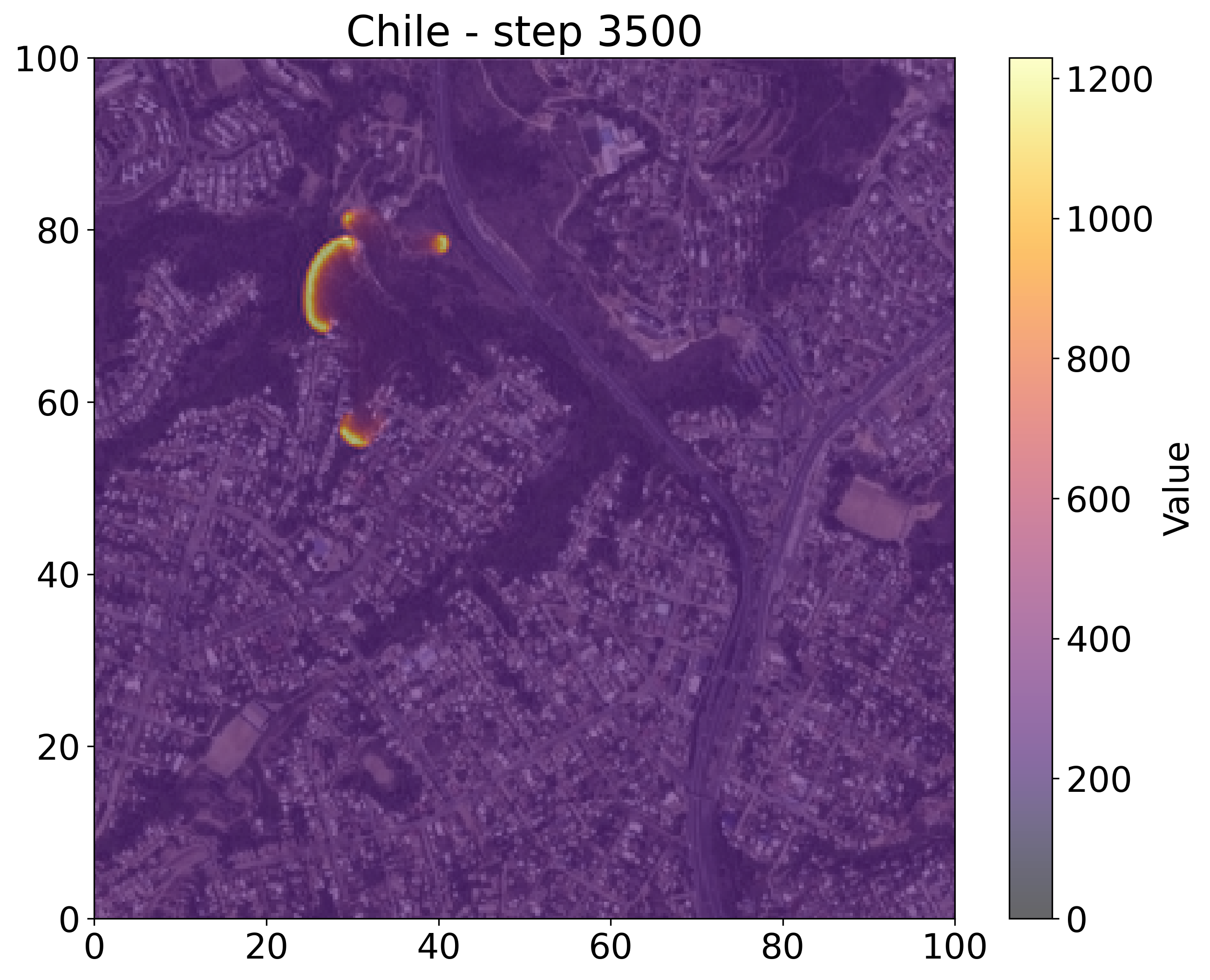}   \includegraphics[width=0.32\textwidth]{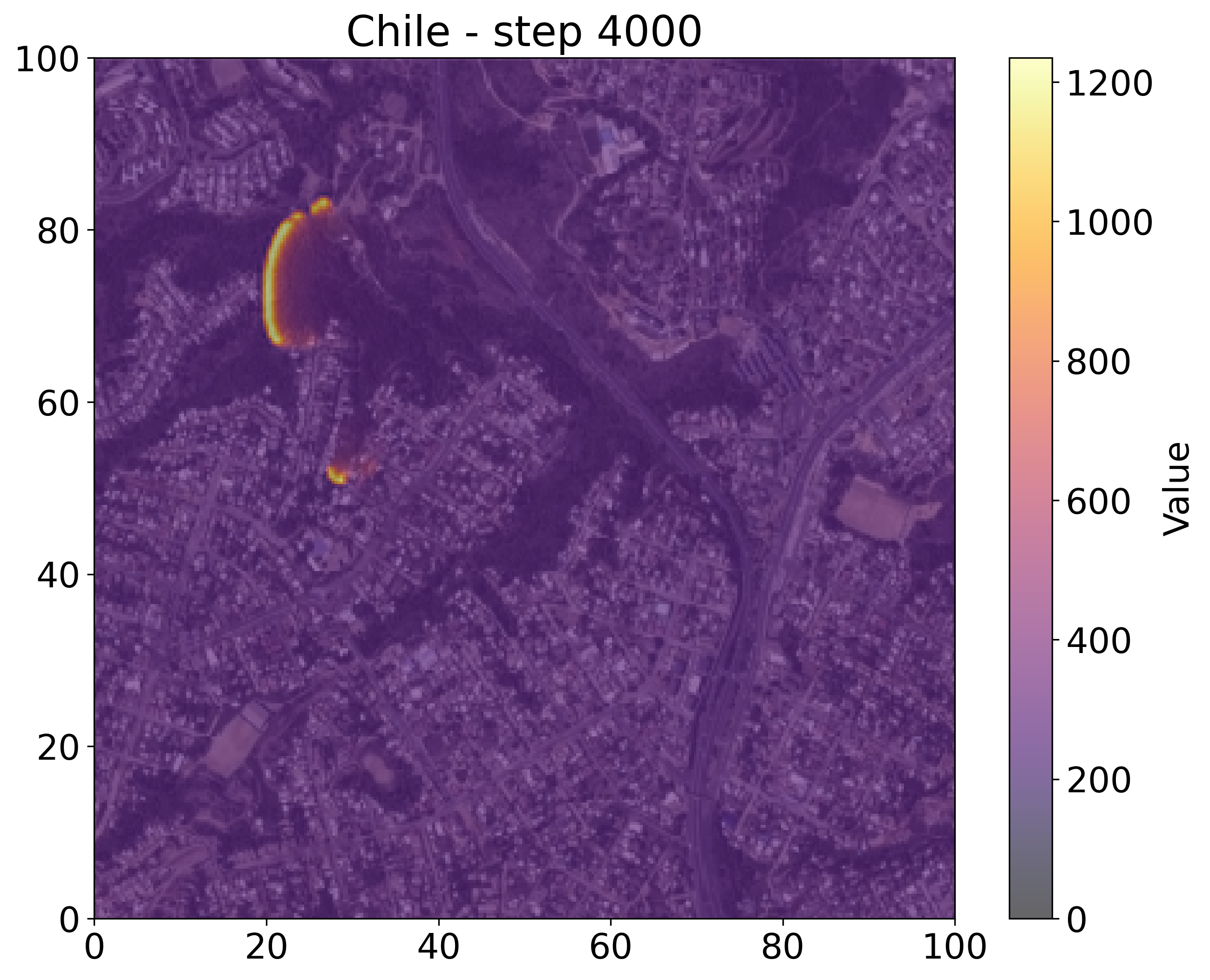} \includegraphics[width=0.32\textwidth]{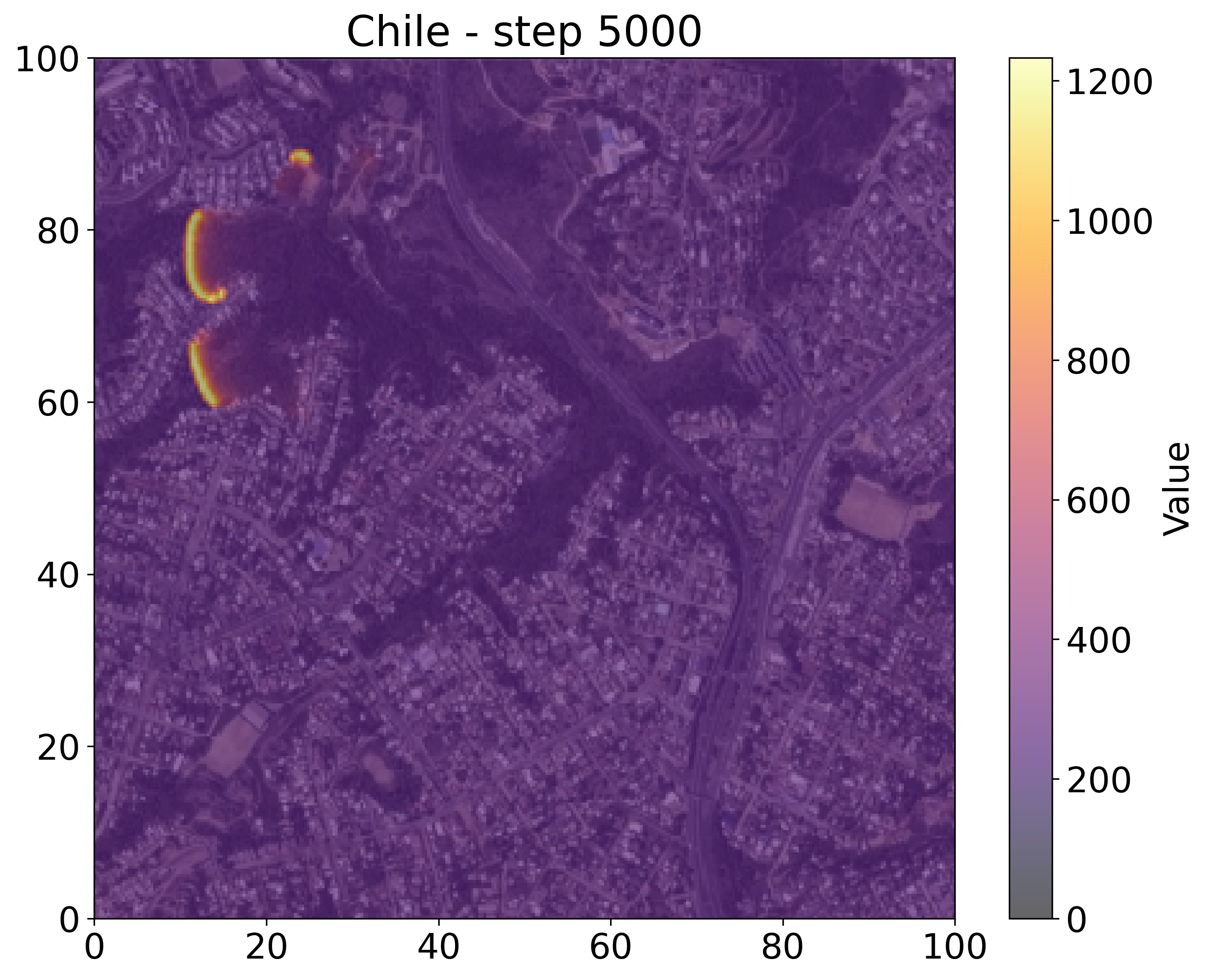}  \\

  \caption{Snapshoots from the wildifre simulation at Viña del Mar.}
  \label{fig:image_grid}
\end{figure}

\subsection{Wildfire of the Las Palmas de Gran Canaria}

\begin{figure}[h]
    \centering
    \includegraphics[width=0.6\textwidth]{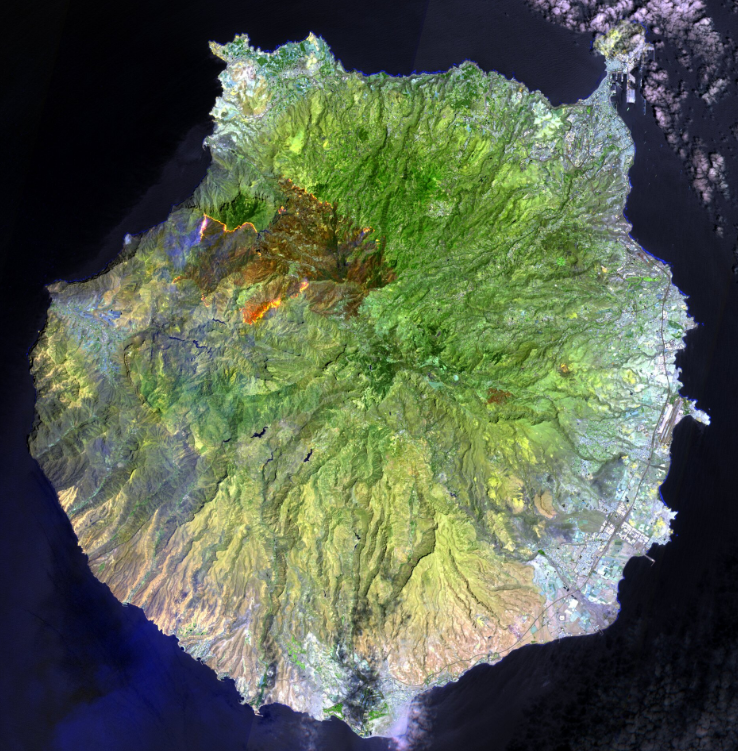}
    \caption{A satellite image taken at the end of the wildfire event.}
    \label{fig:cannaries_real}
\end{figure}

\begin{figure}[h!]
\centering
  \includegraphics[width=0.5\textwidth]{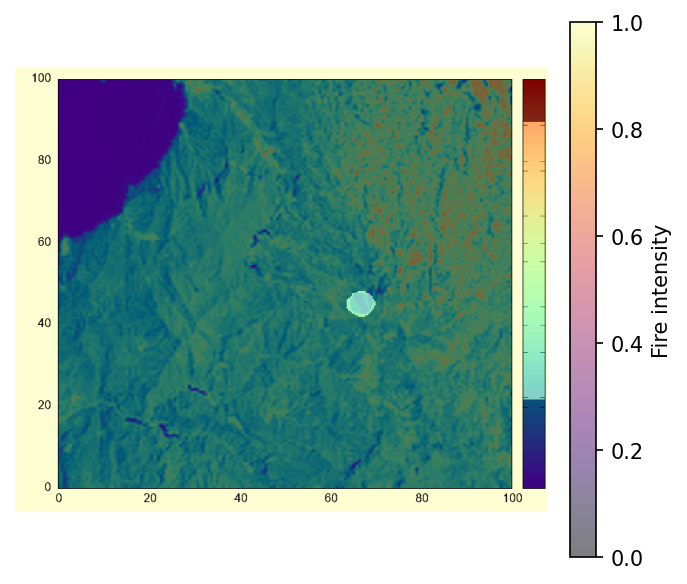} 
  \caption{Initial fire configuration for the Las Palmas de Gran Canaria simulation.}
  \label{fig:initiallas}
\end{figure}

\begin{figure}
\centering
\includegraphics[width=0.32\textwidth]{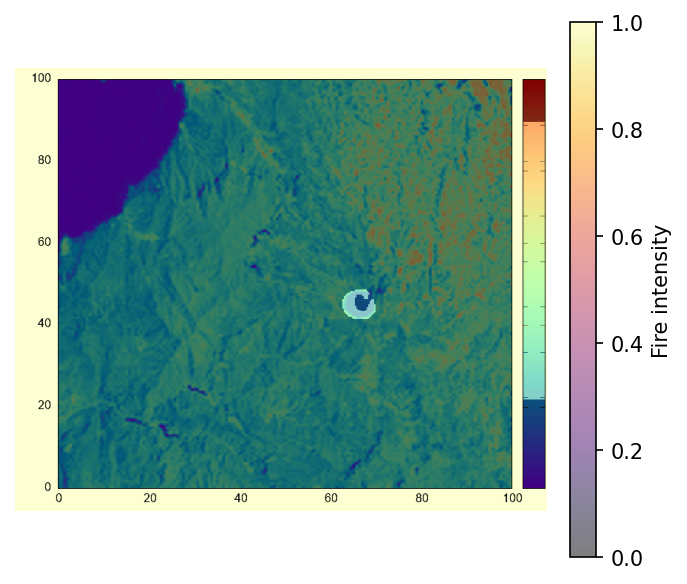}   \includegraphics[width=0.32\textwidth]{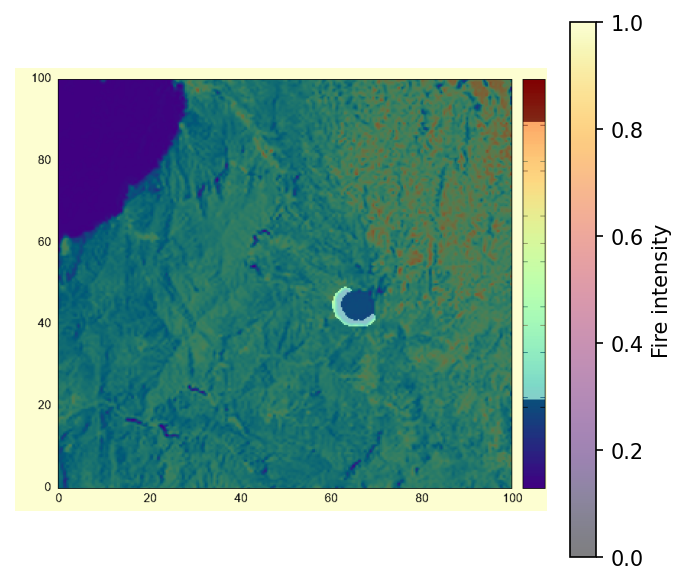} \includegraphics[width=0.32\textwidth]{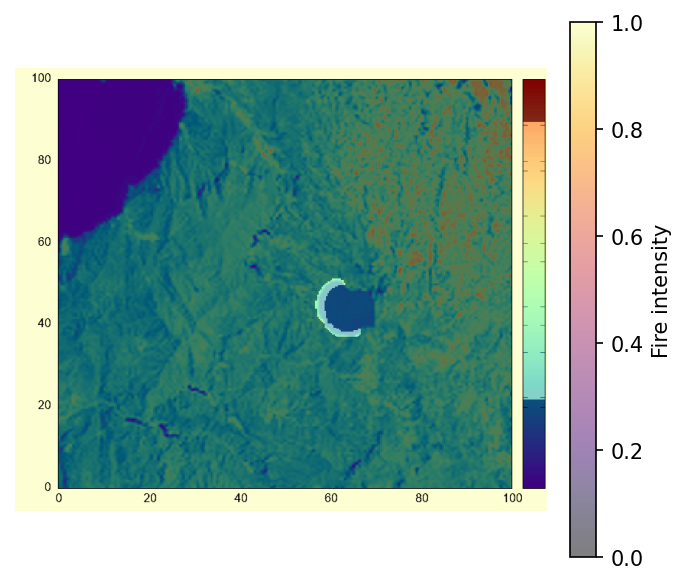}  \\
\includegraphics[width=0.32\textwidth]{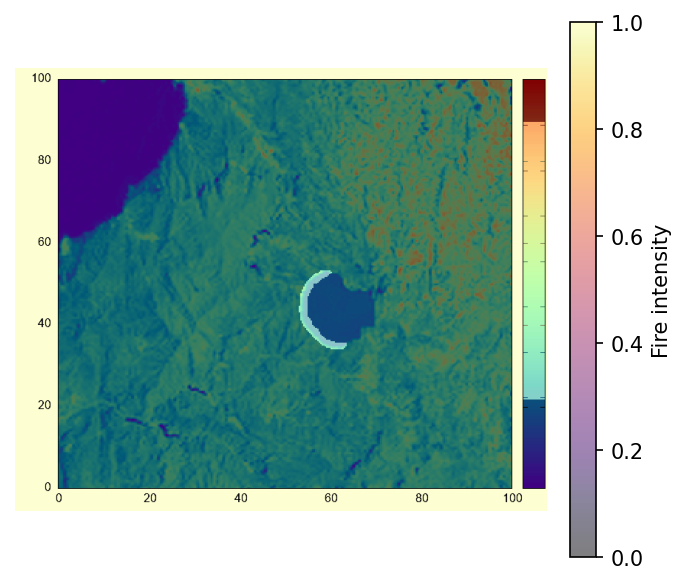}   \includegraphics[width=0.32\textwidth]{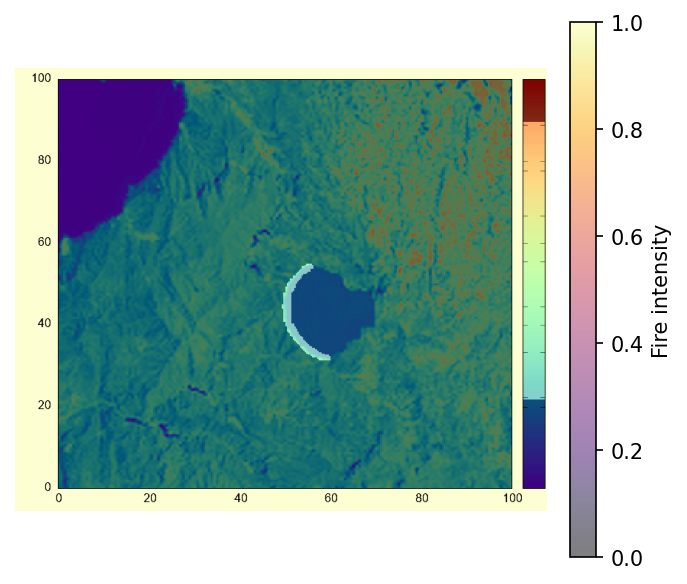} \includegraphics[width=0.32\textwidth]{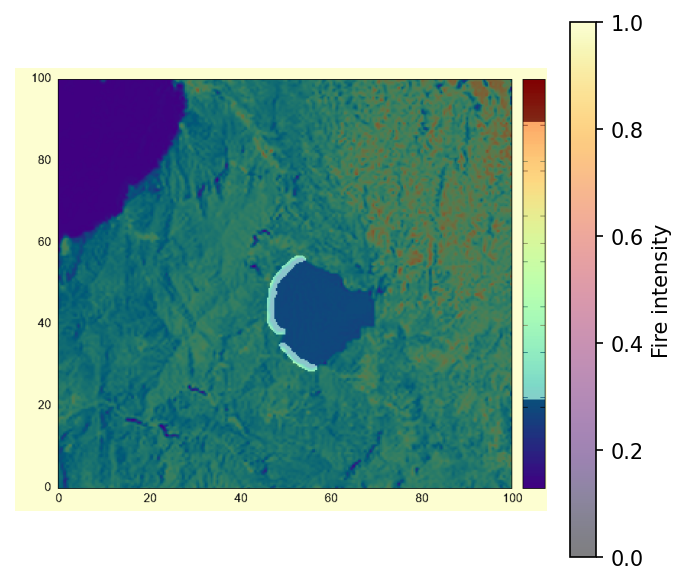}  \\
\includegraphics[width=0.32\textwidth]{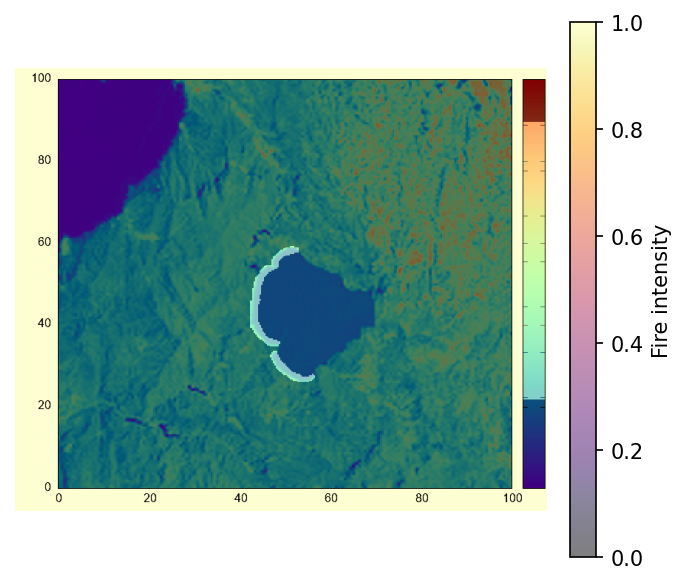}   \includegraphics[width=0.32\textwidth]{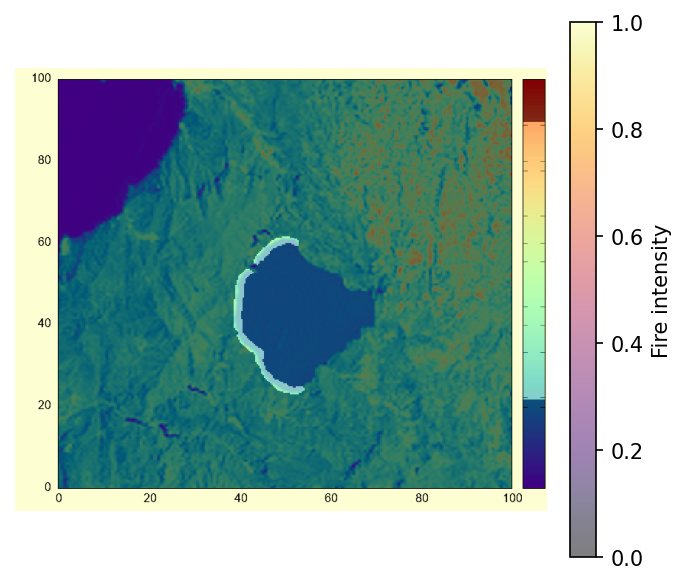} \includegraphics[width=0.32\textwidth]{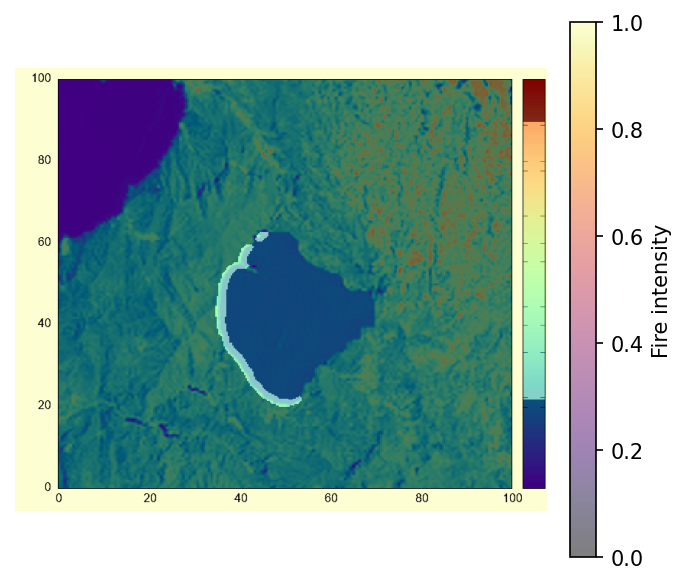}  \\
\includegraphics[width=0.32\textwidth]{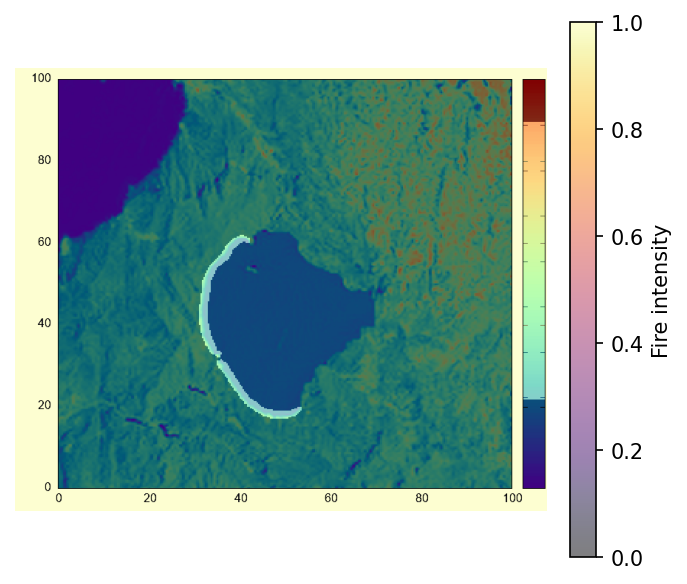}   \includegraphics[width=0.32\textwidth]{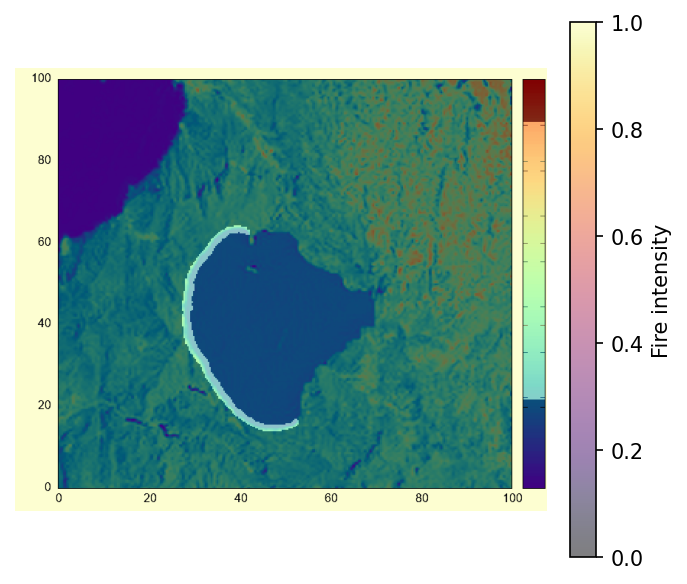} \includegraphics[width=0.32\textwidth]{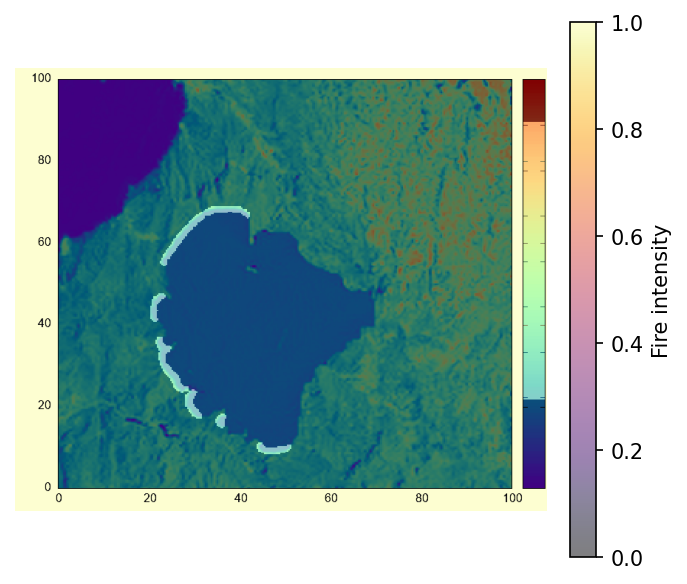}  \\

  \caption{Snapshots from wildfire simulation at Las Palmas de Gran Canaria.}
  \label{fig:image_grid1}
\end{figure}

  We start by focusing on the satellite imaging of Las Palmas de Gran Canaria at the end of the wildfire.
These images are available at \cite{ESACanaries}.
In our simulation, the fuel map, presented in Figure \ref{fig:initiallas}, has been obtained from satellite imagery shown in Figure \ref{fig:cannaries_real}.
We assumed a single ignition point as denoted by a small blue circle in Figure \ref{fig:initiallas}.
The wind direction has been set based on the available two measurement points.
For the first part of the simulation, it is assumed to be the Western-Southern wind. For the second part of the simulation, the wind is from the West, slightly directed to the North.
With these assumptions, we have run the simulations.
The snapshots from the simulation are shown in Figure \ref{fig:image_grid1}, which illustrate the updates to the fuel map and the progress of the wildfire. Additionally, in Figure \ref{fig:image_grid1} we present the snapshots of the simulation illustrating the wildfire front.
When comparing the snapshots from Figures \ref{fig:image_grid1} with the satellite images. Both the real and the simulated wildfires propagated in the form of a triangle.
In the case of real wildfire, there were over 700 firefighters working hard to contain the disastrous phenomenon \cite{ESACanaries} using thirteen planes and a helicopter \cite{ElPais}. They maintained to contain the spread of wildfire, in particular in the Northern part where large forest is present, and in some populated areas in the South. Our simulation shows how the wildfire would spread if there were no active firefighter units.

\section{Scalability of the shared-memory parallel code}

In this section we show the paraller scalability of our simulator.
\revision{Since the most time consuming part of the algorithm is the integration of the right-hand-side, we process the loops over finite elements in parallel.}
\begin{lstlisting}[
    language=C++,
    caption={Code.},
    basicstyle=\ttfamily,
    keywordstyle=\color{blue},
    commentstyle=\color{green},
    stringstyle=\color{red},
    frame=single,
    numbers=left,
    numberstyle=\tiny\color{gray},
    stepnumber=1,
    numbersep=5pt,
    backgroundcolor=\color{lightgray!25}
]
// Compute the right-hand side 
// for temperature and fuel equations
void compute_rhs(double t) {
auto& rhs = u; // RHS for temperature         
auto& rhs_fuel = fuel; // RHS for fuel
zero(rhs); // Reset temperature RHS
zero(rhs_fuel); // Reset fuel RHS
executor.for_each(elements(), [&](index_type e) {
 auto U = element_rhs(); // Local RHS for temperature
 auto F = element_rhs(); // Local RHS for fuel 
 double J = jacobian(e);  // Element Jacobian
 for (auto q : quad_points()) {        
  double w = weight(q); // Quadrature weight        
  auto x = point(e, q); // Physical point 
  // Previous temperature                
  value_type u = eval_fun(u_prev, e, q); 
  // Previous fuel
  value_type fuel = eval_fun(fuel_prev, e, q); 
  for (auto a : dofs_on_element(e)) {
   auto aa = dof_global_to_local(e, a);          
   value_type v = eval_basis(e, q, a); // Basis function           
   // Inverse density * heat capacity          
   double inv = 1.0 / (rho * cp); 
   // Reaction and heat transfer terms
   double delta = u.val>Tig && fuel.val > 0.2 ? 1.0 : 0.0;
   double hc = -70; // Enthalpy
   double r = delta * Ar * u.val * std::exp(-Ta / u.val);
   double Rc = - 1e4 * rho * ch * hc * M / M1 * r;
   double Qw = - rho * cw * (bx * u.dx + by * u.dy);
   // Heat conduction  
   double qc = - kappa * grad_dot(u, v); 
   double qr = -4 * sigma * eps * delta_x 
     * std::pow(u.val, 3) * grad_dot(u, v);
   double Qconv = xi * (T0 - u.val); // Convection
   double Qrz = sigma * eps / delta_z 
     * (std::pow(T0, 4) - std::pow(u.val, 4));          
   double val = (Rc + Qw + Qconv + Qrz) * v.val 
     + qc + qd + qr; // Add temperature term to RHS
   U(aa[0], aa[1]) += 
     (u.val * v.val + steps.dt * inv * val) * w * J;          
   // Add fuel contributions to RHS
   double fval = - delta * 3e2 * r * fuel.val * v.val; 
   F(aa[0], aa[1]) += 
     (fuel.val * v.val + steps.dt * fval) * w * J;
  }
 }
// Update global RHS from local contributions
executor.synchronized([&] 
  { update_global_rhs(rhs, U, e); });
executor.synchronized([&] 
  { update_global_rhs(rhs_fuel, F, e); });
});
\end{lstlisting}

The computations have been executed on a Intel Core Ultra 5 245 processor with 14 physical cores and 32 GB of RAM in DDR dual channel mode. The time is measured in seconds. 

\subsection{Strong scaling}

\begin{figure}[h]
    \centering
    \includegraphics[width=0.8\textwidth]{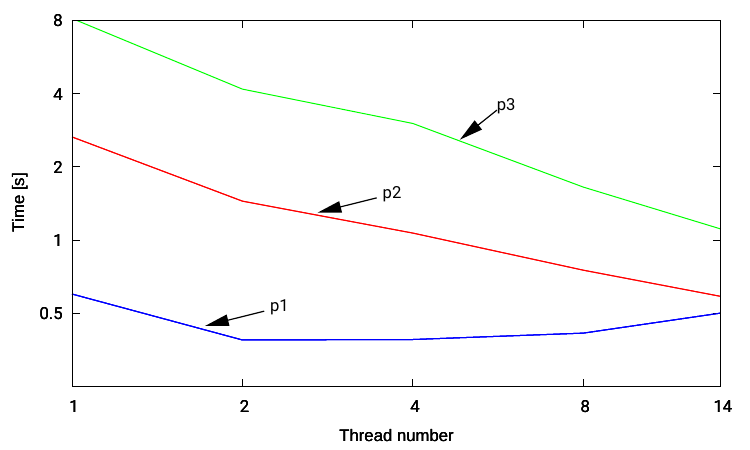}
    \caption{Execution time of 120 time steps of the wildfire simulation over $100\times 100$ computational mesh.}
    \label{fig:time100}
\end{figure}

\begin{figure}[h]
    \centering
    \includegraphics[width=0.8\textwidth]{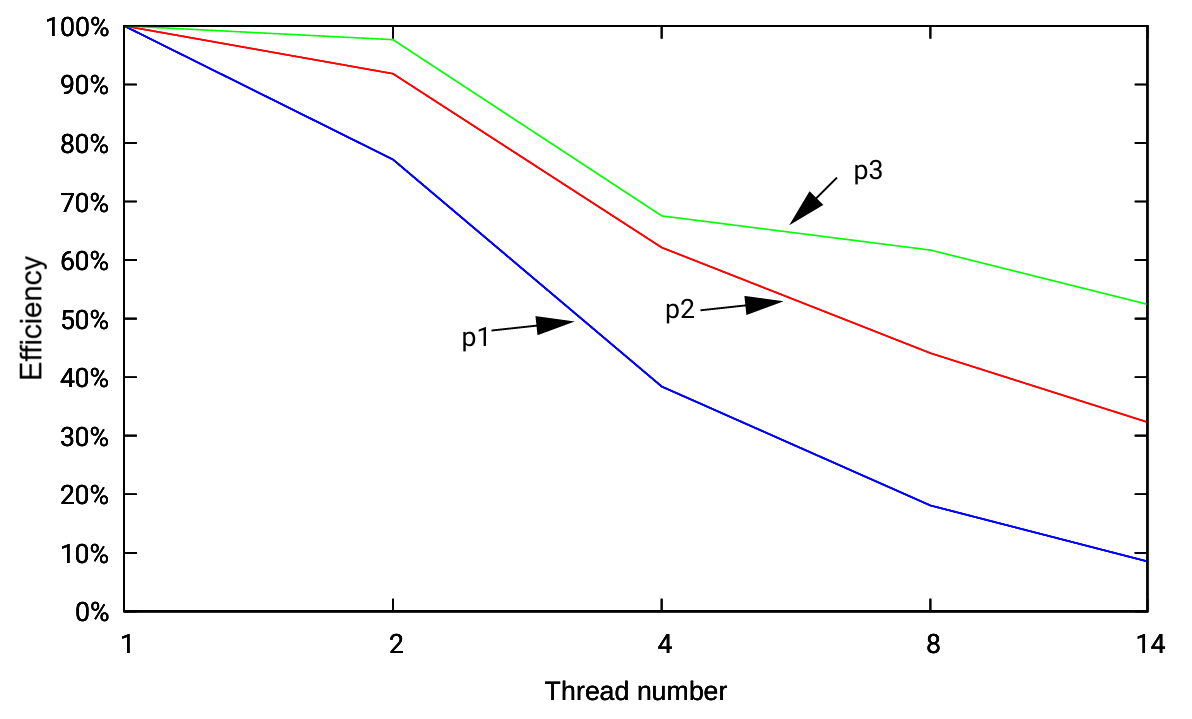}
    \caption{Efficiency of 120 time steps of the wildfire simulation over $100\times 100$ computational mesh.}
    \label{fig:efficiency100}
\end{figure}

\begin{figure}[h]
    \centering
    \includegraphics[width=0.8\textwidth]{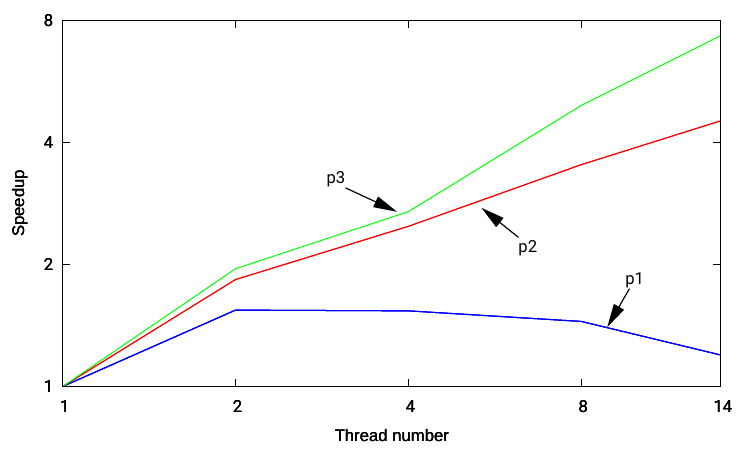}
    \caption{Speedup of 120 time steps of the wildfire simulation over $100\times 100$ computational mesh.}
    \label{fig:speedup100}
\end{figure}

We first tested the execution time of the wildfire simulation code on the computational mesh of size $100 \times 100$ using 120 time steps, 
varying the polynomial order of approximation from $p=1,2,3$, in Figures \ref{fig:time100}-\ref{fig:speedup100}. 
As can be seen from Figure \ref{fig:speedup100}, the speedup reaches 1.5 for linear B-splines, 4.5 for quadratic B-splines, and 7 for cubic B-splines.
Similar behavior can be observed in the efficiency estimates in Figure \ref{fig:efficiency100}.
From the time measurements in Figure \ref{fig:time100}, we see that we can solve the problem over $100\times 100$ mesh in less than 0.5 second for linear B-splines, around 0.6 seconds using quadratic B-splines, and 1.2 seconds using cubic B-splines.

\begin{figure}[h]
    \centering
    \includegraphics[width=0.8\textwidth]{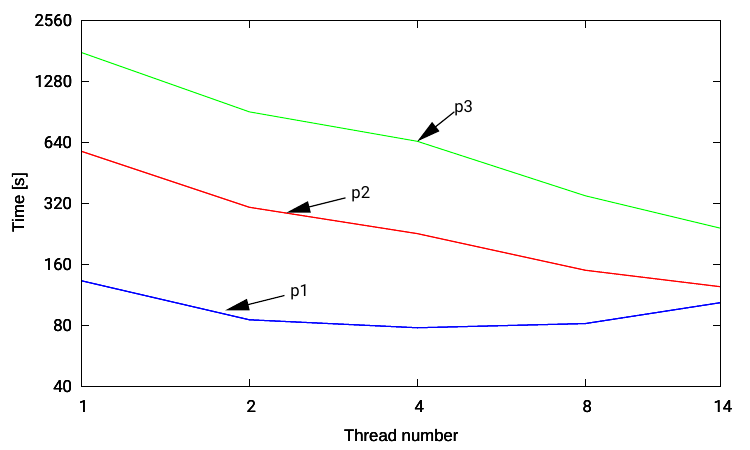}
    \caption{Execution time of 720 time steps of the wildfire simulation over $600 \times 600$ computational mesh.}
    \label{fig:time600}
\end{figure}

\begin{figure}[h]
    \centering
    \includegraphics[width=0.8\textwidth]{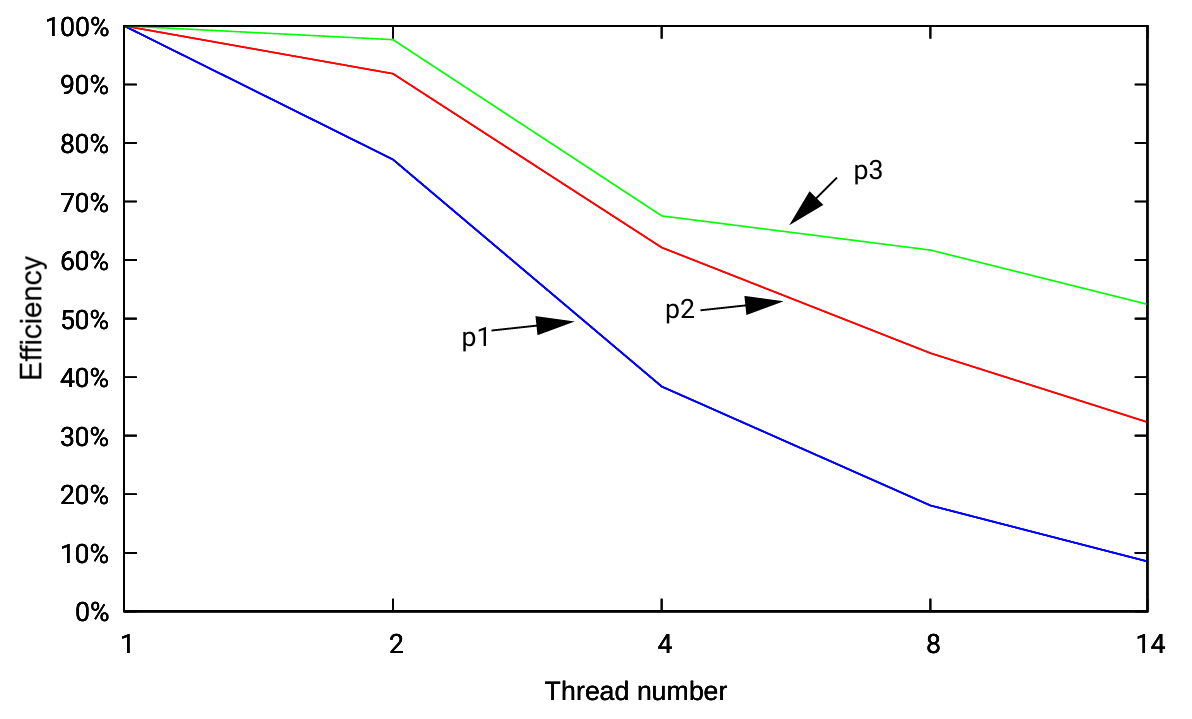}
    \caption{Efficiency of 720 time steps of the wildfire simulation over $600\times 600$ computational mesh.}
    \label{fig:efficiency600}
\end{figure}

\begin{figure}[h]
    \centering
    \includegraphics[width=0.8\textwidth]{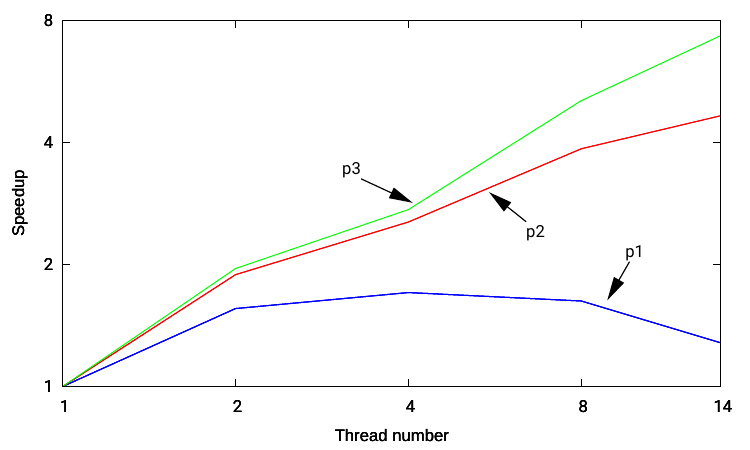}
    \caption{Speedup of 720 time steps of the wildfire simulation over $600\times 600$ computational mesh.}
    \label{fig:speedup600}
\end{figure}

Next, we test the scalability of the wildfire simulations on the computational mesh of size $600 \times 600$ using 720 time steps 
for linear, quadratic, and cubic B-splines.
These experiments are summarized in Figures \ref{fig:time600}-\ref{fig:speedup600}, keeping the ratio between the number of time steps and the mesh size unchanged.
We can see from Figure \ref{fig:speedup600} that the speedup reaches 1.6 for linear B-splines, 5 for quadratic B-splines, and 7.5 for cubic B-splines. 
The efficiency drop when we increase the number of cores is presented in Figure \ref{fig:efficiency600}. We can see that multi-threading is actually doing a good job for linear and quadratic B-splines (since the curve does not turn downward), but it is reducing the efficiency of the cubic B-splines.
The execution time for the problem size on the $600\times 600$ mesh can be read from Figure \ref{fig:time600}.
We can solve the problem in around 80 seconds using linear B-splines, in around 120 seconds using quadratic B-splines, and in around 240 seconds using cubic B-splines.

\begin{figure}[h]
    \centering
    \includegraphics[width=0.8\textwidth]{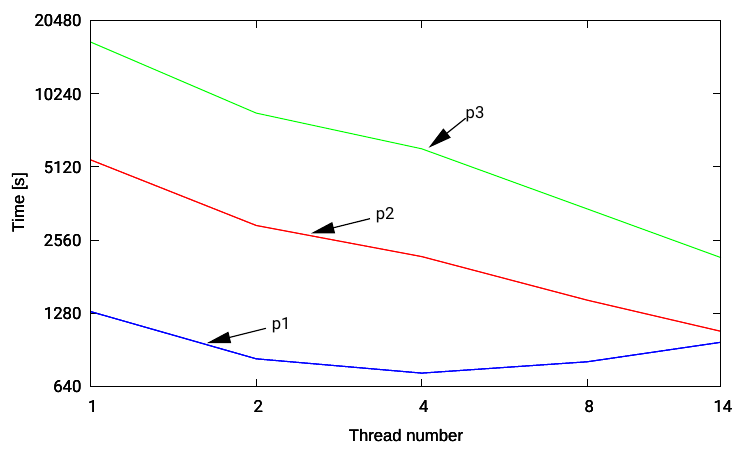}
    \caption{Execution time of 1080 time steps of the wildfire simulation over $1500 \times 1500$ computational mesh.}
    \label{fig:time1500}
\end{figure}

\begin{figure}[h]
    \centering
    \includegraphics[width=0.8\textwidth]{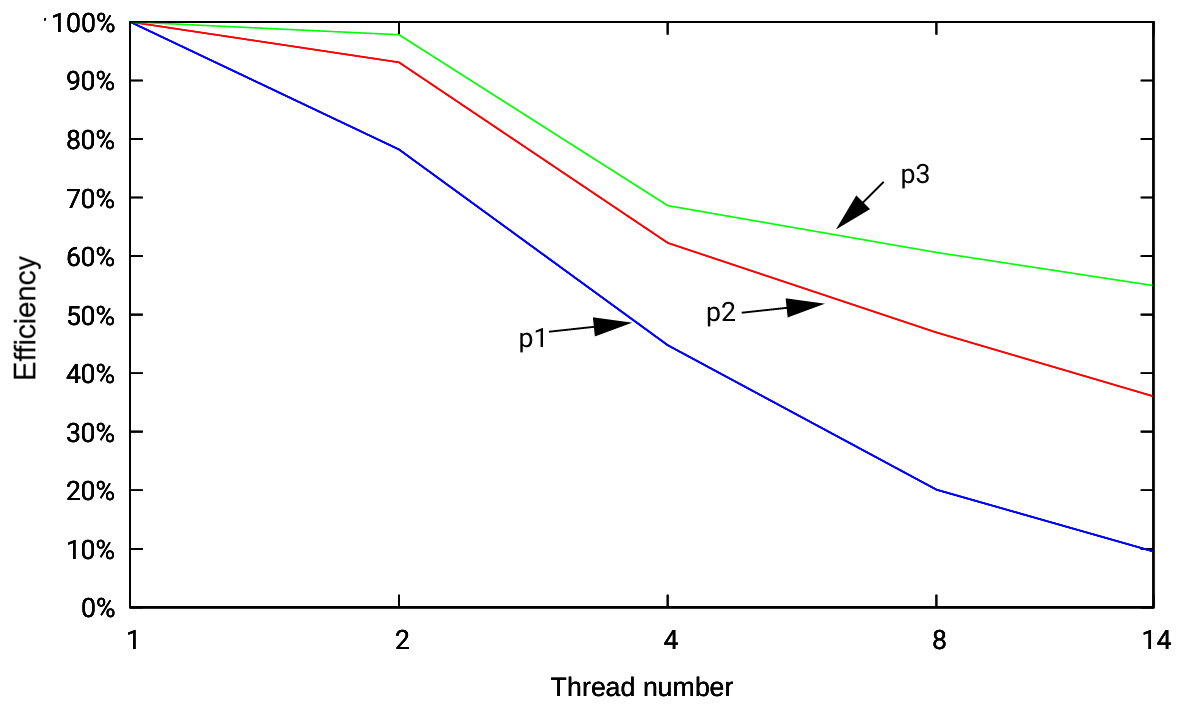}
    \caption{Efficiency of 1080 time steps of the wildfire simulation over $1500\times 1500$ computational mesh.}
    \label{fig:efficiency1500}
\end{figure}

\begin{figure}[h]
    \centering
    \includegraphics[width=0.8\textwidth]{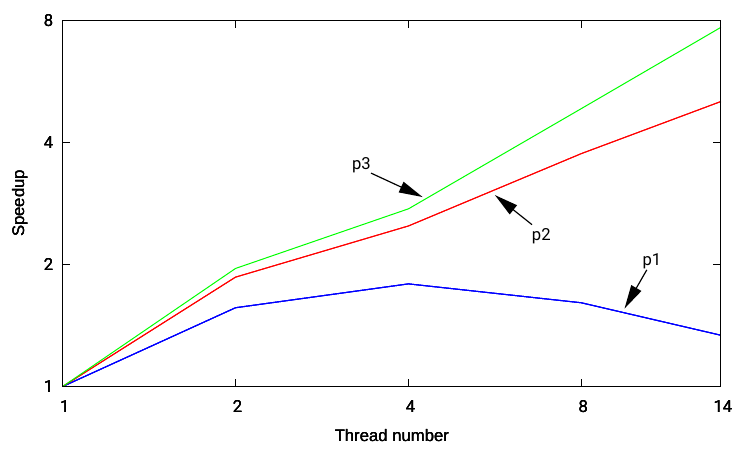}
    \caption{Speedup of 1080 time steps of the wildfire simulation over $1500\times 1500$ computational mesh.}
    \label{fig:speedup1500}
\end{figure}

Finally, we test the scalability of the wildfire simulator on $1500 \times 1500$ mesh, using 1080 time steps (keeping the ratio between the number of time steps and the mesh size constant).
We vary the polynomial order of approximation from $p=1$ through $p=2$, up to $p=3$, as shown in Figures \ref{fig:time1500}-\ref{fig:speedup1500}. 
This time, the speedup `es 1.8 for linear B-splines, 5 for quadratic B-splines, and 7.5 for cubic B-splines.
The efficiency is highest for cubic B-splines, moderate for quadratic B-splines, and lowest for linear B-splines; see Figure \ref{fig:efficiency1500}.
Execution time analysis of Figure \ref{fig:time1500} shows that we can solve the problem over $1500\times 1500$ mesh in around 730 seconds with linear B-splines, around 1080 seconds using quadratic B-splines, and around 2170 seconds using cubic B-splines.

\subsection{Weak scaling}

\begin{figure}[h]
    \centering
    \includegraphics[width=0.8\textwidth]{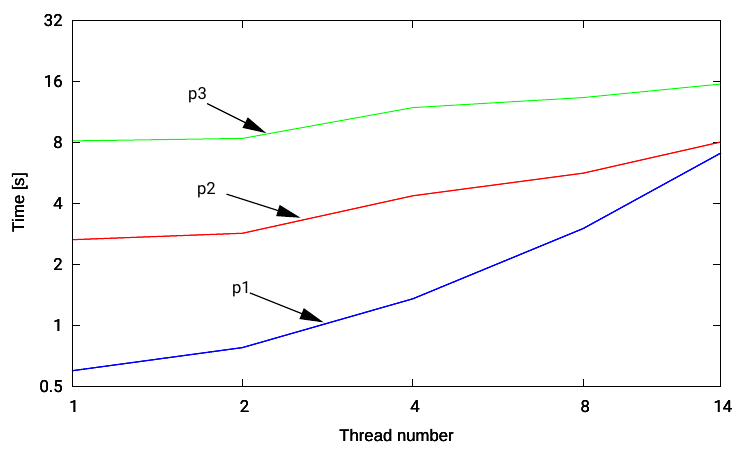}
    \caption{Weak scalability for patches of $100 \times 100$ elements.}
    \label{fig:weak100}
\end{figure}

\begin{figure}[h]
    \centering
    \includegraphics[width=0.8\textwidth]{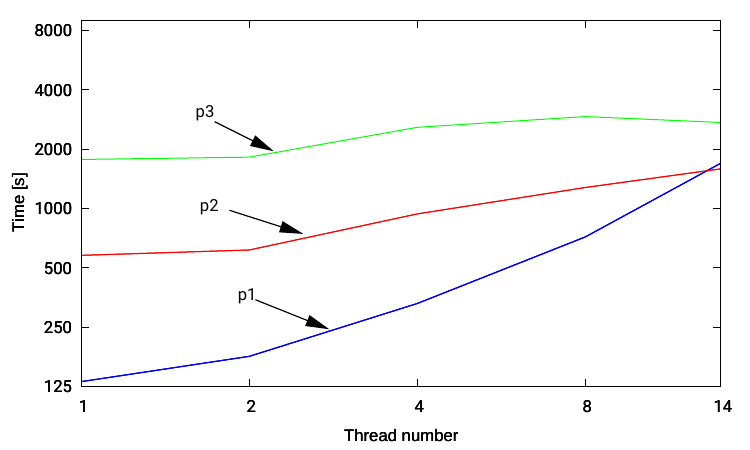}
    \caption{Weak scalability for patches of $600 \times 600$ elements.}
    \label{fig:weak600}
\end{figure}

\begin{figure}[h]
    \centering
    \includegraphics[width=0.8\textwidth]{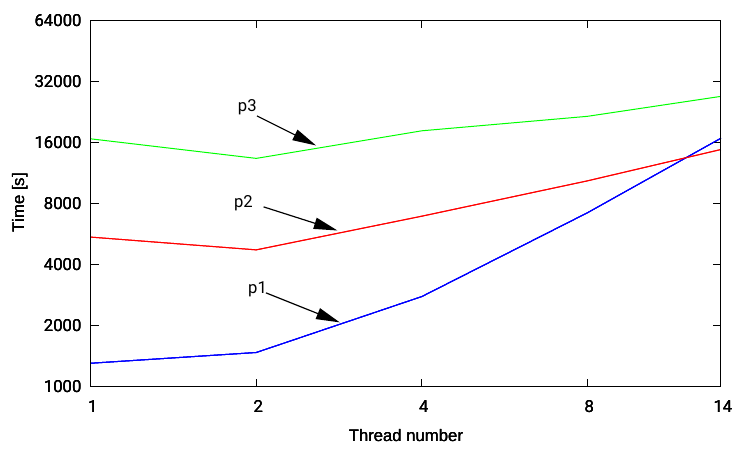}
    \caption{Weak scalability for patches of $1500 \times 1500$ elements.}
    \label{fig:weak1500}
\end{figure}

The weak scaling analysis is presented in Figures \ref{fig:weak100}-\ref{fig:weak1500}. First, we assign a patch of $100\times 100$ elements to a single core and increase the number of cores from 1 to 14 while simultaneously increasing the mesh size. The mesh sizes for the increasing number of cores are summarized in Table \ref{tab:weak100}.
The weak scalability for $100\times 100$ element patches is presented in Figure \ref{fig:weak100}.
For linear B-splines, the execution time grows from 0.6 seconds to 7 seconds for 14 cores.
For quadratic B-splines, the time grows from 2.6 seconds for 1 patch to 8 seconds for 14 cores (mesh size $200 \times 700$).
For cubic B-splines, the single patch mesh takes 8.2 seconds, the 14 patch mesh takes 16 seconds.

The weak scalability for $600\times 600$ element patches is presented in Figure \ref{fig:weak600}.
In this case, the single patch takes 132 seconds for linear B-splines, 578 seconds for quadratic B-splines, and 1775 seconds for cubic B-splines.
This execution time does not change much if we use two patches ($1200\times 600$ elements).
Then, it grows to around 1700 seconds for linear B-splines, 1600 seconds for quadratic B-splines, and 2750 seconds for cubic B-splines with 14 patches (mesh size of $1200\times 1800$ elements).

Finally, the weak scalability for $1500\times 1500$ element patches is presented in Figure \ref{fig:weak1500}.
Here, single patch takes 1300 seconds for linear B-splines, 5500 seconds for quadratic B-splines, and 17000 seconds for cubic B-splines.
The execution time decreases when we move to two patches (the mesh size $3000 \times 1500$ elements), down to around 4800 seconds for quadratic B-splines, and 13500 seconds for cubic B-splines.
We do not know how to explain this phenomenon. Next, the execution time increases to around 17000 seconds for linear B-splines, 15000 seconds for quadratic B-splines, and 2700 seconds for cubic B-splines when considering 14 cores and $3000\times 10500$ elements.

\begin{table}[h]
    \centering
    \begin{tabular}{c|c|c|c}
        Number  & Mesh for & Mesh for & Mesh for  \\ 
        of cores & patch $100 \times 100$ & patch  $600 \times 600$ & patch $1500 \times 1500$  \\         \hline 
        1 & $100 \times 100$ &  $600 \times 600$ &   $1500 \times 1500$ \\
        2 & $200 \times 100$ & $1200 \times 600$ &   $3000 \times 1500$  \\
        4 & $200 \times 200$ &  $1200 \times 1200$ &  $3000 \times 3000$ \\
        8 & $400 \times 200$ &  $2400 \times 1200$ &  $6000 \times 3000$ \\
        14 & $200 \times 700$ &  $1200 \times 4200$ &   $3000 \times 10500$\\
    \end{tabular}
    \caption{Mesh sizes for increasing number of cores, for different patch sizes.}
    \label{tab:weak100}
\end{table}

\section{Conclusions}

We presented a quasi-implicit linear-computational-cost numerical scheme and an isogeometric analysis solver for running wildfire simulations. We analyzed the scheme and method experimentally. We tested the solver's convergence and accuracy on the model problem with a manufactured solution.
We employed the wildfire simulator to run two numerical simulations.
The first example concerned the wildfire that happened at the beginning of 2024 in central and southern Chile. From February 1 to 5, 2024, a series of wildfires occurred.
For the sake of simulation, we chose a specific urban region of Viña del Mar.
The second simulation concerned the wildfire that happened on the island of Las Palmas in Gran Canaria between August 17 and 20, 2019. We compared both simulations to satellite data and available reports.
We provided a source code repository with brief instructions for setting up and running new wildfire simulations.
Future work will involve developing the Variational Physics-Informed Neural Network model \cite{kharazmi2021hp}, including discrete weak formulations and the collocation method for CRVPINN \cite{LOS2025107839}.

\section*{Acknowledgments}

The authors are grateful for the support from the funds that the Polish Ministry of Science and Higher Education assigned to AGH University of Krakow.
The work is supported by the ``Excellence initiative - research university" for AGH University of Krakow.

\section*{Declaration of Generative AI and AI-assisted technologies in the writing process}
During the preparation of this work, the author(s) used Gemini (Google)  and ChatGPT to assist with writing, correct English language usage, and improve text clarity and organization. After using this tool/service, the author(s) reviewed and edited the content as needed and take(s) full responsibility for the content of the publication.

 \bibliographystyle{elsarticle-num} 
 \bibliography{cas-refs-no-url}

\end{document}